\documentclass[aps,twocolumn,prc,superscriptaddress,showpacs,nofootinbib]{revtex4}
\usepackage{graphicx}
\usepackage{dcolumn}
\usepackage{bm}
\usepackage{color}
\usepackage{amsmath}    
\usepackage{graphicx}   

\begin{document}

\title{Fission barriers in covariant density functional theory:
extrapolation to superheavy nuclei.}
%
\author{H. Abusara}
\affiliation{Department of Physics and Astronomy, Mississippi State
University, MS 39762}
\affiliation{Department of Electronics and Applied Physics, 
Palestine Polytechnic University, Hebron, Palestinian territory}
\author{A. V. Afanasjev}
\affiliation{Department of Physics and Astronomy, Mississippi State
University, MS 39762}
\author{P.\ Ring}
\affiliation{Fakult\"at f\"ur Physik, Technische Universit\"at M\"unchen,
 D-85748 Garching, Germany}

\date{\today}

\begin{abstract}
Systematic calculations of fission barriers allowing for triaxial
deformation are performed for even-even superheavy nuclei with charge
number $Z=112-120$ using three classes of covariant density functional
models. The softness of nuclei in the triaxial plane leads to an
emergence of several competing fission pathes in the region of the inner
fission barrier in some of these nuclei. The outer fission barriers are
considerably affected by triaxiality and octupole deformation. General 
trends of the evolution of the inner and the outer fission barrier heights 
are discussed as a function of the particle numbers.
\end{abstract}

\maketitle

\section{Introduction}

  The on-going search for new superheavy elements is motivated by the attempts
to provide theoretical and experimental explanations for two open questions in
nuclear structure. The first question is related to the limits of the existence
of atomic nuclei at large values of proton number, while the second one to the
location of the island of stability of superheavy nuclei and the next magic
numbers (if any) beyond $Z=82$ and $N=126$.  Heavy and superheavy nuclei decay
by spontaneous fission and therefore their stability is defined essentially by
the size and the shape of the fission barriers. Thus, the inner fission barrier
is considered a fundamental characteristic of these nuclei which is important
in resolving the questions mentioned above.

  A systematic investigation of the properties of fission barriers is the
best way to address the stability of nuclei against spontaneous fission
since it eliminates the arbitrariness of conclusions with respect to the
choice of a specific nucleus. Different models with triaxiality included
have been used for extensive theoretical investigations of the properties
of inner fission barriers in the actinides.
It has been found that the heights of inner fission barriers are reduced
when triaxial shapes are taken into account
\cite{AAR.10,SBDN.09,WERP.02}. However, this reduction has a strong dependence
on the particle number in the proton and neutron subsystems as well as on the
applied model. These investigations were performed in the following frameworks:
microscopic+macroscopic (MM) methods \cite{MSI.04,SK.06,DPB.07,MSI.09,DNPB.09,KJS.10},
the extended Thomas-Fermi plus Strutinsky integral (ETFSI) method \cite{DPT.00},
and non-relativistic energy density functionals (EDF) based on zero-range Skyrme
\cite{SBDN.09,BRRMG.98,BQS.04,SDN.06} and finite-range Gogny
\cite{WERP.02,DGGL.06,WAR.09,RSRG.10} forces, and recently in covariant density
functional theory (CDFT) \cite{AAR.10,LNVR.10,RAALNV.11}.

  These methods have also been used for the study of fission barriers
in superheavy nuclei (see Table \ref{tab4} below for a review).
However, only few of them take into account triaxiality. For example,
within the framework of covariant density functional theory,
it has been considered only in the study of fission
barriers in a single nucleus ($^{264}$Hs) in Ref.\ \cite{BRRMG.98}. Thus,
in order to fill this gap in our knowledge a systematic study of
fission barriers of even-even $Z=112-120$ superheavy nuclei with
triaxiality included is performed in the current manuscript.  We use
the same method that has been successfully employed
in Ref.\ \cite{AAR.10} for the study of inner fission barriers in actinides,
where an average deviation from experiment of 0.76 MeV has been found.
When extrapolating to superheavy nuclei we want to understand the impact of
triaxiality on fission barriers and how they evolve with the change of the
particle numbers. An additional goal  is to see how the fission barriers
depend on the choice of the specific CDFT model. While the systematic
investigation of inner fission barriers of actinides in Ref.\
\cite{AAR.10} has been performed with the nonlinear meson-nucleon coupling
model represented by the NL3* parametrization of the relativistic mean
field (RMF) Lagrangian, in the current study we use in addition to
the nonlinear coupling model  also density-dependent meson-nucleon and
point coupling models.

 Density functional theories are extremely useful for the microscopic description
of quantum mechanical many-body systems. They have been applied with
great success for many years in Coulombic systems~\cite{KS.65,KS.65a}, where they
are in principle exact and where the functional can be derived without any
phenomenological adjustments directly from the Coulomb interaction. In nuclear
physics with spin and isospin degrees of freedom, with strong nucleon-nucleon and
three-body forces the situation is much more complicated. However, covariant density
functionals exploit
basic properties of QCD at low energies, in particular symmetries and the separation
of scales~\cite{LNP.641}. They provide a consistent treatment of the spin degrees of
freedom, they include the complicated interplay between the large Lorentz scalar and
vector self-energies induced on the QCD level by the in-medium changes of the scalar
and vector quark condensates~\cite{CFG.92}. In addition, these
functionals include {\it nuclear magnetism} \cite{KR.89}, i.e. a consistent
description of currents and time-odd mean fields important for odd-mass nuclei \cite{AA.10},
the excitations with unsaturated spins, magnetic moments \cite{HR.88} and nuclear
rotations \cite{AR.00,TO-rot}. Because of Lorentz invariance no new adjustable parameters
are required for the time-odd parts of the mean fields. Of course, at present,
all attempts to derive these functionals directly from the bare
forces~\cite{BT.92,HKL.01,SOA.05,HSR.07}
do not reach the required accuracy. However, in recent years
modern and very successful covariant density
functionals have been derived~\cite{DD-ME1,DD-ME2,DD-PC1,DD-ME3}
which are based on density dependent vertices and one additional parameter
characterizing the range of the force. They provide an excellent description of
ground states and excited states all over the periodic table~\cite{VALR.05,NVR.11}
with a high predictive power. Modern versions of these forces derive the density
dependence of the vertices from state-of-the-art ab-initio calculations and use
only the remaining few parameters for a fine tuning of experimental masses in finite
spherical~\cite{DD-ME3} and deformed~\cite{DD-PC1} nuclei.

 The manuscript is organized as follows. The theoretical framework and the details
of the numerical calculations are discussed in Sec.\ \ref{Theor-sect}. The results
of the investigations of the fission barriers, the role of triaxiality  and the
comparison with the actinide region are presented in Sec.\ \ref{res+diss}.
Finally, Sec.\ \ref{conc} summarizes the results of our work.

\section{Theoretical framework and the details of numerical calculations}
\label{Theor-sect}

  Three classes of covariant density functional models are used throughout
this manuscript: the nonlinear meson-nucleon coupling model (NL), the
density-dependent meson-exchange model (DD-ME) and a density-dependent point
coupling model (DD-PC). The main differences between them lay in the treatment of
the range of the interaction, the mesons  and density dependence.
The interaction in the first two classes has a finite range,
while the third class uses zero-range interaction with one
additional gradient term in the scalar-isoscalar channel. The mesons are
absent in the density-dependent point coupling model. The density dependence is
explicit in the last two models, while it shows up via the non-linearity in the
$\sigma$-meson in the nonlinear meson-nucleon coupling model. Each of
these classes is represented in the current manuscript by a set of parameters
that is considered to be state-of-the-art.

 In the meson-exchange models \cite{NL3*,TW.99,DD-ME2}, the nucleus is described
as a system of Dirac nucleons interacting via the exchange of mesons with finite
masses leading to finite-range interactions. The starting point of covariant
density functional theory (CDFT) for these two models is a standard Lagrangian
density~\cite{GRT.90}
\begin{align}
\mathcal{L}  &  =\bar{\psi}\left[
\gamma(i\partial-g_{\omega}\omega-g_{\rho
}\vec{\rho}\,\vec{\tau}-eA)-m-g_{\sigma}\sigma\right]  \psi\nonumber\\
&  +\frac{1}{2}(\partial\sigma)^{2}-\frac{1}{2}m_{\sigma}^{2}\sigma^{2}%
-\frac{1}{4}\Omega_{\mu\nu}\Omega^{\mu\nu}+\frac{1}{2}m_{\omega}^{2}\omega
^{2}\label{lagrangian}\\
&  -\frac{1}{4}{\vec{R}}_{\mu\nu}{\vec{R}}^{\mu\nu}+\frac{1}{2}m_{\rho}%
^{2}\vec{\rho}^{\,2}-\frac{1}{4}F_{\mu\nu}F^{\mu\nu}\nonumber
\end{align}
which contains nucleons described by the Dirac spinors $\psi$ with
the mass $m$ and several effective mesons characterized by the
quantum numbers of spin, parity, and isospin. They create effective
fields in a Dirac equation, which corresponds to the Kohn-Sham
equation~\cite{KS.65} in the non-relativistic case.
The Lagrangian (\ref{lagrangian}) contains as parameters the meson
masses $m_{\sigma}$, $m_{\omega}$, and $m_{\rho}$ and the coupling
constants $g_{\sigma}$, $g_{\omega}$, and $g_{\rho}$. $e$ is the
charge of the protons and it vanishes for neutrons. This linear
model has first been introduced by Walecka~\cite{Wal.74,SW.86}.

 To treat the density dependence in this model Boguta and Bodmer~\cite{BB.77}
introduced a density dependence via a non-linear meson coupling
replacing the term $\frac{1}{2}m_{\sigma}^{2}\sigma^{2}$ in Eq.
(\ref{lagrangian}) by
\begin{equation}
U(\sigma)~=~\frac{1}{2}m_{\sigma}^{2}\sigma^{2}+\frac{1}{3}g_{2}\sigma
^{3}+\frac{1}{4}g_{3}\sigma^{4}.
\end{equation}
The nonlinear meson-nucleon coupling is represented by the parameter
set NL3* \cite{NL3*} (see Table \ref{tab1}), which is a
modern version of the widely used parameter set NL3 \cite{NL3}. Apart
from the fixed values for the masses $m$, $m_\omega$ and $m_\rho$, there
are six phenomenological parameters $m_\sigma$, $g_\sigma$, $g_\omega$,
$g_\rho$, $g_2$, and $g_3$.

 The density-dependent meson-nucleon coupling model has an explicit density
dependence for the meson-nucleon vertices.
There are no nonlinear terms in the $\sigma$
meson, i.e. $g_2 = g_3 =0$.
The meson-nucleon vertices are defined as:
\begin{equation}
 g_{i}(\rho) = g_i(\rho_{\rm sat})f_i(x) \quad {\rm for} \quad i=\sigma, \omega, \rho
\end{equation}
where the density dependence is given by
\begin{equation}\label{fx}
 f_i(x)=a_i\frac{1+b_i(x+d_i)^2}{1+c_i(x+d_i)^2}.
\end{equation}
for $\sigma$ and $\omega$ and by
\begin{equation}
 f_\rho(x)=\exp(-a_\rho(x-1)).
\end{equation}
for the $\rho$ meson. \textit{x} is defined as the ratio between the baryonic
density $\rho$ at a specific location and the baryonic density at saturation
$\rho_{\rm sat}$ in symmetric nuclear matter. The eight parameters in Eq.\ (\ref{fx})
are not independent, but constrained as follows:
$f_i(1)=1$, $f_{\sigma}^{''}(1)=f_{\omega}^{''}(1)$, and $f_{i}^{''}(0)=0$. These
constraints reduce the number of independent parameters for density dependence to
three. This model is represented in the present investigations by the parameter
set DD-ME2 \cite{DD-ME2} given in Table \ref{tab1}.

\begin{table}[ptb]
\caption{ The parameters of the NL3* and DD-ME2 parameterizations
of the Lagrangian. Note that
$g_{\sigma}=g_{\sigma}(\rho_{\rm sat})$,
$g_{\omega}=g_{\omega}(\rho_{\rm sat})$ and
$g_{\rho}=g_{\rho}(\rho_{\rm sat})$
in the case of the DD-ME2 parametrization.
\label{tab1}}
\begin{center}%
\begin{tabular}
[c]{|c|c|c|}\hline Parameter &  NL3* &  DD-ME2 \\\hline
$m$ &  939  & 939  \\
$m_{\sigma}$ &  502.5742  & 550.1238 \\
$m_{\omega}$ &  782.600  & 783.000  \\
$m_{\rho}$ &  763.000  & 763.000 \\
$g_{\sigma}$ &  10.0944 &  10.5396\\
$g_{\omega}$ &  12.8065 & 13.0189\\
$g_{\rho}$ &  4.5748 & 3.6836  \\
$g_{2}$ &  -10.8093  &         \\
$g_{3}$ &  -30.1486  &         \\
$a_{\sigma}$ &       & 1.3881 \\
$b_{\sigma}$ &       & 1.0943 \\
$c_{\sigma}$ &       & 1.7057 \\
$d_{\sigma}$ &       & 0.4421 \\
$a_{\omega}$ &       & 1.3892 \\
$b_{\omega}$ &       & 0.9240 \\
$c_{\omega}$ &       & 1.4620 \\
$d_{\omega}$ &       & 0.4775 \\
$a_{\rho}$   &       & 0.5647 \\
\hline
\end{tabular}
\end{center}
\end{table}

 The Lagrangian for the density-dependent point coupling model \cite{NHM.92,DD-PC1}
is given by
\begin{align}
\mathcal{L}  &  =\bar{\psi}\left(i\gamma \cdot \partial-m\right)  \psi\nonumber\\
&  -\frac{1}{2}\alpha_S(\hat\rho)\left(\bar{\psi}\psi\right)\left(\bar{\psi}\psi\right)
-\frac{1}{2}\alpha_V(\hat\rho)\left(\bar{\psi}\gamma^{\mu}\psi\right)\left(\bar{\psi}\gamma_{\mu}\psi\right)\nonumber\\
&-\frac{1}{2}\alpha_{TV}(\hat\rho)\left(\bar{\psi}\vec\tau\gamma^{\mu}\psi\right)\left(\bar{\psi}\vec\tau\gamma_{\mu}\psi\right)\nonumber\\
&-\frac{1}{2}\delta_S\left(\partial_v\bar{\psi}\psi\right)\left(\partial^v\bar{\psi}\psi\right) - e\bar\psi\gamma \cdot A
\frac{(1 - \tau_3)}{2}\psi .
\label{Lag-pc}
\end{align}
It contains the free-nucleon Lagrangian, the point coupling interaction terms,
and the coupling of the proton to the electromagnetic field. The derivative terms
in Eq.\ (\ref{Lag-pc}) account for the leading effects of finite-range interaction
which are important in nuclei. In analogy with meson-exchange models, this model
contains isoscalar-scalar, isoscalar-vector and isovector-vector interactions.
In the present work it is represented by the DD-PC1 parametrization \cite{DD-PC1}
given in Table \ref{tab2}.

\begin{table}[ptb]
\caption{The parameters of the DD-PC1 parameterization in the RMF Lagrangian}%
\label{tab2}
\begin{center}%
\begin{tabular}
[c]{|c|c|}\hline Parameter &  DD-PC1 \\\hline
$m$ &   939\\
$a_{\sigma}$ & -10.04616\\
$b_{\sigma}$ & -9.15042\\
$c_{\sigma}$ & -6.42729\\
$d_{\sigma}$ &  1.37235\\
$a_{\omega}$ &  5.91946\\
$b_{\omega}$ &  8.86370\\
$d_{\omega}$ &  0.65835\\
$b_{\rho}$   &  1.83595\\
$d_{\rho}$   &  0.64025\\
\hline
\end{tabular}
\end{center}
\end{table}

\begin{table}[ht]
\caption{The $G_1^n$, $G_2^n$, $G_1^p$ and $G_2^p$ parameters [in
MeV] for different parameterizations of the RMF Lagrangian.}
\label{tab3}
\renewcommand{\tabcolsep}{0.5pc}
\renewcommand{\arraystretch}{1.4}
\begin{tabular} {ccccc}\hline \hline
Force  & $G_1^n$ &  $G_2^n$  &  $G_1^p$ & $G_2^p$ \\ \hline
NL3*   &  10.7    &   -10.4  &   7.40   & 18.9   \\
DD-PC1 &  10.5    &   -7.38  &   7.5    & 19.2   \\
DD-ME2 &  11.0    &   -10.3  &   7.9    & 17.0   \\ \hline \hline
\end{tabular}
\\[2pt]\end{table}

 The triaxial relativistic mean field (RMF)+BCS approach \cite{KR.88} is
used here for the description of fission barriers. This approach has been
very successfully applied to a systematic description of the fission
barriers in the actinides \cite{AAR.10}. The RMF-equations are solved
iteratively and at each iteration the BCS occupation probabilities
$v_{k}^{2}$ are determined. These quantities are used in the calculation of densities,
energies and new fields for the next iteration. We use a monopole
pairing force \cite{RS.80} with the strength parameters $G_{\tau}$ for
neutrons ($\tau=n$) and protons ($\tau=p$).

  For each type of particle we start with a pairing strength
parameter $G$ and solve at each iteration the gap equation~\cite{RS.80}
\begin{equation}
\frac{1}{G}=%
\sum\limits_{k>0}\frac{1}{2E_{k}}%
\label{gap-equation}%
\end{equation}
with $E_{k}=\sqrt{(\varepsilon_{k}-\lambda)^{2}+\Delta^{2}}$,  where
$\varepsilon_{k}$ are the eigenvalues of the Dirac equation and
the chemical potential $\lambda$ is determined by the average
particle number. Then the occupation probabilities
\begin{equation}
v_{k}^2=\frac{1}{2}\left(  1-\frac{\varepsilon_{k}-\lambda}{E_{k}%
}\right),  \label{v2}%
\end{equation}
and the gap parameters
\begin{equation}
\Delta=G\sum\limits_{k>0}u_{k}v_{k}%
\label{delta-BCS}%
\end{equation}
are determined in a self-consistent way. The pairing energy is defined
as
\begin{equation}
E_{\rm pair}=-\Delta%
\sum\limits_{k>0}
u_{k}v_{k}, %
\label{Epair}%
\end{equation}
The sum over $k$ in Eqs.~(\ref{gap-equation}), (\ref{delta-BCS}) and
(\ref{Epair}) run over all states in the pairing window $E_k<E_{\rm cutoff}$.
 In Ref.\ \cite{MN.92} empirical pairing gap parameters
\begin{eqnarray}
\Delta_{n}^{\rm emp} = \frac{4.8}{N^{1/3}}\,\,\,\, {\rm MeV}, \qquad
\Delta_{p}^{\rm emp} = \frac{4.8}{Z^{1/3}}\,\,\,\, {\rm MeV}
\label{emp-G}
\end{eqnarray}
have been determined by a fit to experimental
data on neutron and proton gaps in the normal deformed minimum.

\begin{figure}[ht]
\includegraphics[angle=0,width=8.0cm]{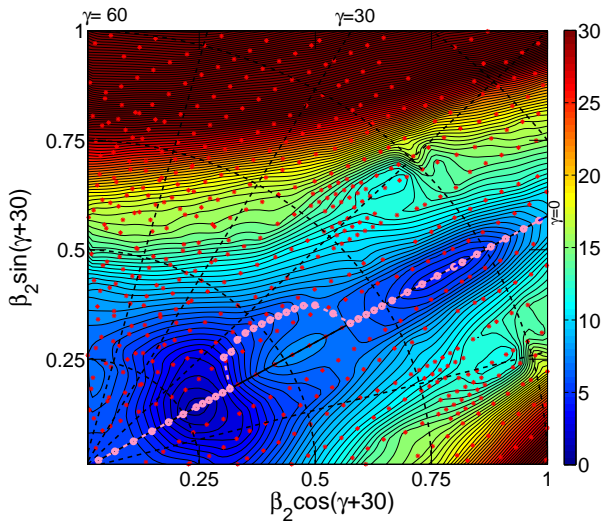}
\caption{(Color online) Potential energy surface of the $^{240}$Pu
nucleus with the NL3* parametrization of the RMF Lagrangian. The
energy difference between two neighboring equipotential lines is
equal to 0.5 MeV. The pink dashed line with solid circles shows the
lowest-energy solution as a function of $\beta_2$. Small red stars
show the positions of deformation points at which the numerical
results were obtained. The figure is based on the results of
Ref.\ \cite{AAR.10}. Further details are given in the text.}
\label{pes-2D}
\end{figure}

\begin{figure*}[ht]
\begin{center}
\includegraphics[width=12.0cm,angle=0]{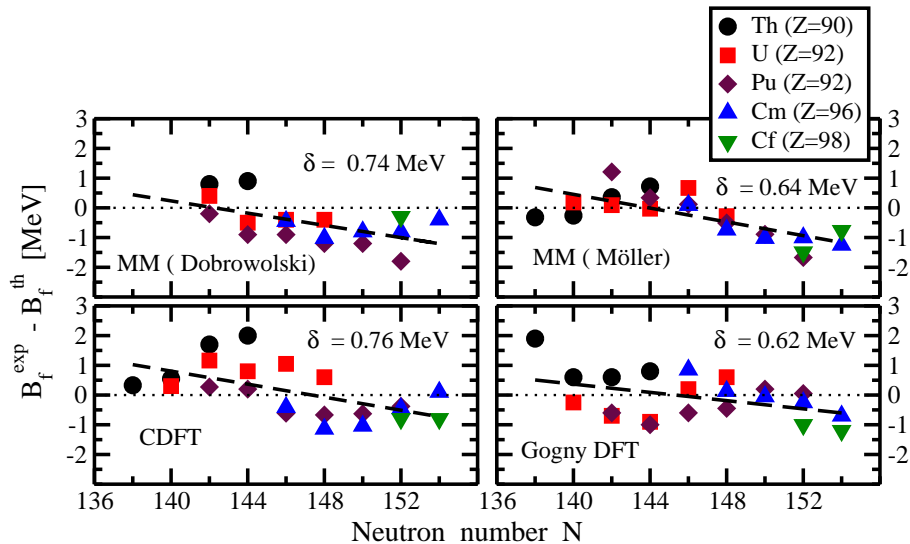}
\end{center}
\caption{(Color online) The difference between experimental and calculated
heights of inner fission barriers as a function of neutron number $N$. The
results of the calculations are compared to estimated fission barrier
heights given in the RIPL-2 database \cite{RIPL-2}, which is used
for this purpose in the absolute majority of theoretical studies on fission
barriers in actinides. The results of the calculations within microscopic+macroscopic
method ('MM(Dobrowolski)' \cite{DPB.07} and 'MM(M{\"o}ller)' \cite{MSI.09}), covariant
density functional theory ('CDFT' \cite{AAR.10}) and density functional
theory based on the finite range Gogny force ('Gogny DFT' \cite{DGGL.06})
are shown. Thick dashed lines are used to show the average trend of the
deviations between theory and experiment as a function of neutron number.
The average deviation per barrier $\delta$ [in MeV] is defined as 
$\delta = \sum_{i=1}^N |B_f^i(th)-B_f^i(exp)|/N$, where $N$ is the number 
of the barriers with known experimental heights, and $B_f^i(th)$ 
($B_f^i(exp)$) are calculated (experimental) heights of the barriers.
Long-dashed lines represent the trend of the deviations between theory and
experiment as a function of neutron number. They are obtained via linear
regression based on a least square fit.}
\label{N-dep}
\end{figure*}

\begin{figure*}[ht]
\begin{center}
\includegraphics[width=14.0cm,angle=0]{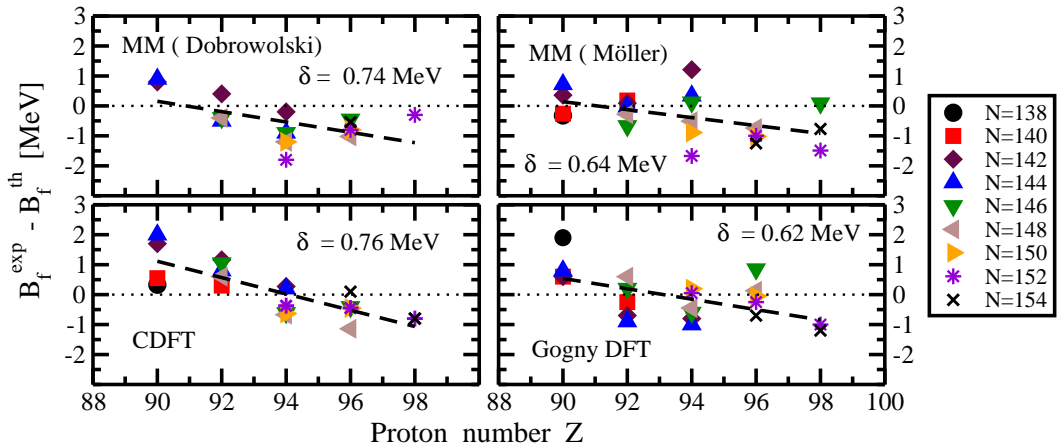}
\end{center}
\caption{ (Color online) The same as in Fig.\ \ref{N-dep} but
as a function of proton number $Z$.}
\label{Z-dep}
\end{figure*}

  These empirical gap parameters form the basis for the definition of
the strength parameters $G_{\tau}$ in the current manuscript. The
following expressions \cite{DMS.80}
\begin{eqnarray}
A \cdot G_n = G_1^n - G_2^n \frac{N-Z}{A}\,\,\,\,{\rm MeV} \\
A \cdot G_p = G_1^p + G_2^p \frac{N-Z}{A}\,\,\,\,{\rm MeV}
\label{G-strength}
\end{eqnarray}
are used in the calculations. First, using empirical gap parameters
of Eq.\ (\ref{emp-G}), the values $G_n(Z,N)$ and $G_p(Z,N)$ are obtained
for the ground states of all even-even nuclei in the
$Z=112-126$, $N-Z=48-62$ region. Then, the parameters $G_1^n$, $G_2^n$,
$G_1^p$ and $G_2^p$ are defined by the least square fit to the set of
the $G_n(Z,N)$ and $G_p(Z,N)$. Their values depend on the parameter set
of the Lagrangian and they are given in Table \ref{tab3}. In this way
we have strength parameters for the effective pairing interaction
depending in a smooth way on the neutron and proton numbers and,
because of the changing level density, the gap parameters derived
from those values show fluctuations as a function of the particle
numbers. Note that similarly to Ref.\ \cite{AAR.10}, the cutoff energy
for the pairing window is set to E$_{\rm cutoff}$=120 MeV.

\begin{table*}[ht]
\centering
\begin{minipage}{16cm}
\caption{The definition of pairing in the studies of inner fission barriers
within the last decade.  The first column shows the first author and the reference.
The pairing model (BCS or HFB) and type of pairing ($G$ - seniority pairing with
fixed strength $G$, $\delta$ - zero range $\delta$-force, $Gogny$ - finite range Gogny
force) are shown in column 2. Column 3 shows either the region of the nuclear chart or
the nucleus in which the fitting of the parameters of the pairing force has been performed in the
case of constant G and/or $\delta$-pairing. Column 4 shows whether particle number
projection (PNP) by means of the Lipkin-Nogami method has been used in the calculations.
Column 5 indicates whether the systematic calculations, covering all even-even actinide
nuclei with measured inner fission barriers, have been performed (Yes) or not (No). In
the case of restricted calculations, the number of nuclei, for which the
calculations have been performed, is shown. Column 6 is similar to column 5 but for
SHE. The calculations are considered to be systematic if they cover a
significant range of proton and neutron numbers. ``T'' (``A'') letter in column
7 indicates that the triaxial deformation is (is not) included into the
calculations. Column 8 shows the pairing window used in the calculations
with constant $G$ and
$\delta$ pairing; no pairing window is used in the calculations with the Gogny
force.
Note that it is not always possible to extract these details
from the original publications or references quoted therein. In these cases, the
relevant box of the table is empty.}
\renewcommand{\tabcolsep}{0.25pc} 
\renewcommand{\arraystretch}{1.4} 
\begin{tabular}{|c|c|c|c|c|c|c|c|c|}
\hline
Author [reference]  & Pairing Model   &  Fitting region   & PNP & Actinide & SHE
& A/T & $E_{\rm cutoff}$      \\ \hline
1    &  2  & 3 & 4 & 5 & 6 & 7 & 8 \\ \hline
%
\multicolumn{8}{|c|}{\bf Macroscopic+microscopic method} \\ \hline
%
%
M\"{o}ller 2009 \cite{MSI.09} & BCS($G$)\cite{MN.92} &  & Yes & Yes & Yes & T & \\ \hline
Dobrowolski 2007 \cite{DPB.07} &  BCS($G$) &  & No & Yes & 2 & T &\\ \hline
Kowal 2010 \cite{KJS.10} &  BCS($G$) & $Z\geq 84$ \cite{MPS.01} & No & Yes & Yes &  T & \\ \hline
%
\multicolumn{8}{|c|}{\bf Extended Thomas-Fermi plus Strutinsky integral} \\ \hline
%
Dutta 2000 \cite{DPT.00} & BCS($\delta$) &  & No & 5 & 5 &  T & \\ \hline
%
\multicolumn{8}{|c|}{\bf Skyrme density functional theory} \\ \hline
%
  Bonneau 2004 \cite{BQS.04} & BCS($G$)/BCS($\delta$) &
  $^{254}No$/$A\sim 178$ &
  No & Yes & No & A/T\footnote{The calculations with allowance
   for triaxial deformation have been performed only for four nuclei}
   & 6 MeV
\\ \hline
  B\"{u}rvenich 2004 \cite{BBM.04} & BCS($\delta$) & across nuclear  &
  No  & Yes &  Yes & A  &  \cite{BRRM.00}\footnote{The
  pairing-active space $\Omega_q$ is chosen to
  include approximately one additional oscillator shell of states above the Fermi level.}
 \\
   & & chart \cite{BRRM.00} & & & & &  \\ \hline
  Samyn 2005 \cite{SGP.05} & HFB($\delta$)  & $Z\in(92,98)$\footnote{Pairing
  force fitted to absolute masses. As a result, it is considerably stronger than
  the one fitted to even-odd mass differences \cite{SGP.05}.}
  & Yes & Yes & Yes & T &  different $E_{\rm cutoff}$\footnote{The single-particle
  states in
  the pairing window $\epsilon_F \pm E_{\rm cutoff}$ are included. $E_{\rm cutoff}=17$ MeV
  for the BSk6, BSk7 and BSk8 Skyrme forces, $E_{\rm cutoff}=15$ MeV for BSk2 force,
  $E_{\rm cutoff}=16.5$ MeV for BSk9 force, $E_{\rm cutoff}=5$ MeV for SLy6 force.}
  \\ \hline
  Staszczak 2006 \cite{SDN.06} & BCS(G) & $^{252}Fm$ and Ref.\cite{MNK.97} & No & Yes & Yes & T &
  lowest Z(N) states\footnote{The pairing-active space consisted of the lowest Z(N) proton
  (neutron) single-particle states.} \\ \hline
 Staszczak 2007 \cite{SDN.07} & BCS($G$)/BCS($\delta$) & $^{252}Fm$ & No & Yes & Yes & T &
 lowest Z(N) states$^{e}$ \\ \hline
%
\multicolumn{8}{|c|}{\bf Gogny density functional theory} \\ \hline
%
Warda 2002 \cite{WERP.02} & HFB(Gogny) & No  & No & 5  & No & T & No \\ \hline
Delaroche 2006 \cite{DGGL.06}  & HFB(Gogny) & No  & No & Yes & No & T & No \\ \hline
%
\multicolumn{8}{|c|}{\bf Covariant density functional theory} \\ \hline
%
Bender 1998 \cite{BRRMG.98}& BCS($\delta$) & across nuclear & No & No & 3\footnote{The
   calculations with allowance for triaxial deformation have been performed only for
   a single nucleus}  & T & \cite{BRRM.00}$^{b}$  \\
 & & chart \cite{BRRM.00} & & & & &  \\ \hline
B\"{u}rvenich 2004 \cite{BBM.04} & BCS($\delta$) & across nuclear &
No  & Yes & Yes & A  & \cite{BRRM.00}$^{b}$ \\
 & & chart \cite{BRRM.00} & & & & &  \\ \hline
Karatzikos 2010 \cite{KALR.10}  & RHB(Gogny)\footnote{RHB=Relativistic Hartree Bogoliubov approach} &
 No    & No  & Yes & Yes & A & No \\ \hline
Abusara 2010 \cite{AAR.10} & BCS($G$) & $Z\in(90,100)$   &  No
 & Yes & No  & T & 120 MeV \\
                                 &                & $N-Z\in(42,66)$   &
  &           &           & &             \\ \hline
\end{tabular}\\[2pt]
\label{tab4}
\end{minipage}
\end{table*}

  The calculations are performed by successive diagonalizations using the
Broyden method \cite{BAM.08} and the method of quadratic constraints 
\cite{RS.80}. We have also implemented the augmented Lagrangian method 
\cite{SSBN.10} for constraints in our computer codes, but we did not find 
any clear advantage of this method over the method of quadratic constraints 
for the type of potential energy surfaces (PES) we are dealing with. 
Starting from the three multipole operators 
\begin{eqnarray}
\hat{Q}_{20}&=&2z^2-x^2-y^2\\
\hat{Q}_{22}&=&x^2-y^2\\
\hat{Q}_{30}&=&z(2z^2-3x^2-3y^2)
\end{eqnarray}
we calculate in the following investigations three types of potential 
energy surfaces, using three types of constraints.

  First, we restrict ourselves to axially symmetric configurations
with reflection symmetry, abbreviated throughout the paper by (A). 
Here we use the computer code DIZ~\cite{RGL.97} based on an expansion
of the Dirac spinors in terms of harmonic oscillator wave functions 
with cylindrical symmetry and we minimize 
\begin{eqnarray}
\left<{H}\right> + C_{20} (\langle\hat{Q}_{20}\rangle-q_{20})^2
\end{eqnarray}
where $\left<{H}\right>$ is the total energy, 
$\langle\hat{Q}_{20}\rangle$ denotes the expectation values of 
the mass quadrupole operators, $q_{20}$ is the constrained value 
of the multipole moment, and $C_{20}$ the corresponding stiffness 
constant \cite{RS.80}.

   In the next step abbreviated throughout the paper by (T) we use the 
triaxial computer code DIC~\cite{KR.89} with the D2-symmetry based
on an expansion of the Dirac spinors in terms of a cartesian oscillator basis
and by imposing constraints on axial and triaxial mass quadrupole 
moments we minimize 
\begin{eqnarray}
\left<{H}\right> + \sum_{\mu=0,2} C_{2\mu}
(\langle\hat{Q}_{2\mu}\rangle-q_{2\mu})^2.
\end{eqnarray}

  In addition, starting from the axial reflection symmetric computer code 
DIZ of Ref.\ \cite{RGL.97}, we have developed an axial reflection asymmetric 
[octupole deformed] code DOZ. It is used for a study of the impact of axial 
octupole deformation on the inner and outer fission barriers. The calculations 
in this axially symmetric octupole code, abbreviated throughout this paper  
by (O) vary the function
\begin{eqnarray}
\langle H\rangle + \sum_{\mu=2,3} C_{\mu,0}
(\langle\hat{Q}_{\mu,0}\rangle-q_{\mu,0})^2.
\end{eqnarray}

 It was checked that the numerical results obtained for axially 
symmetric solutions in the A, T and O calculations differ by no more 
than 50 keV for all deformations of interest.

The truncation of the basis is performed in all these calculations in such a way 
that all states belonging to the shells up to $N_{F}=20$ fermionic shells and $N_{B}=20$
bosonic shells are taken into account. This truncation scheme has been
tested and used in the actinides in Ref.\ \cite{AAR.10}.
Although the calculations in such a truncation scheme provide sufficient
numerical accuracy, they are also very computationally demanding (see Ref.\
\cite{AAR.10} for details). This is the reason why we treat the pairing
channel in the present triaxial RMF calculations in the BCS approximation
despite the fact that the triaxial cranked Relativistic Hartree+Bogoliubov
(RHB) approach with the finite range Gogny forces in the pairing channel
was developed in the late nineties~\cite{AKR.99,ARK.00}. RMF+BCS calculations
are less time-consuming and more stable (especially, in the saddle point region)
than the RHB calculations.

 Fig.\ \ref{pes-2D} shows an example of the distribution of deformation
points on a potential energy surface (PES) at which the numerical results
were obtained using these constraints in a triaxial (T) calculation.

\begin{figure*}[ht]
\begin{center}
\includegraphics[width=8.0cm,angle=0]{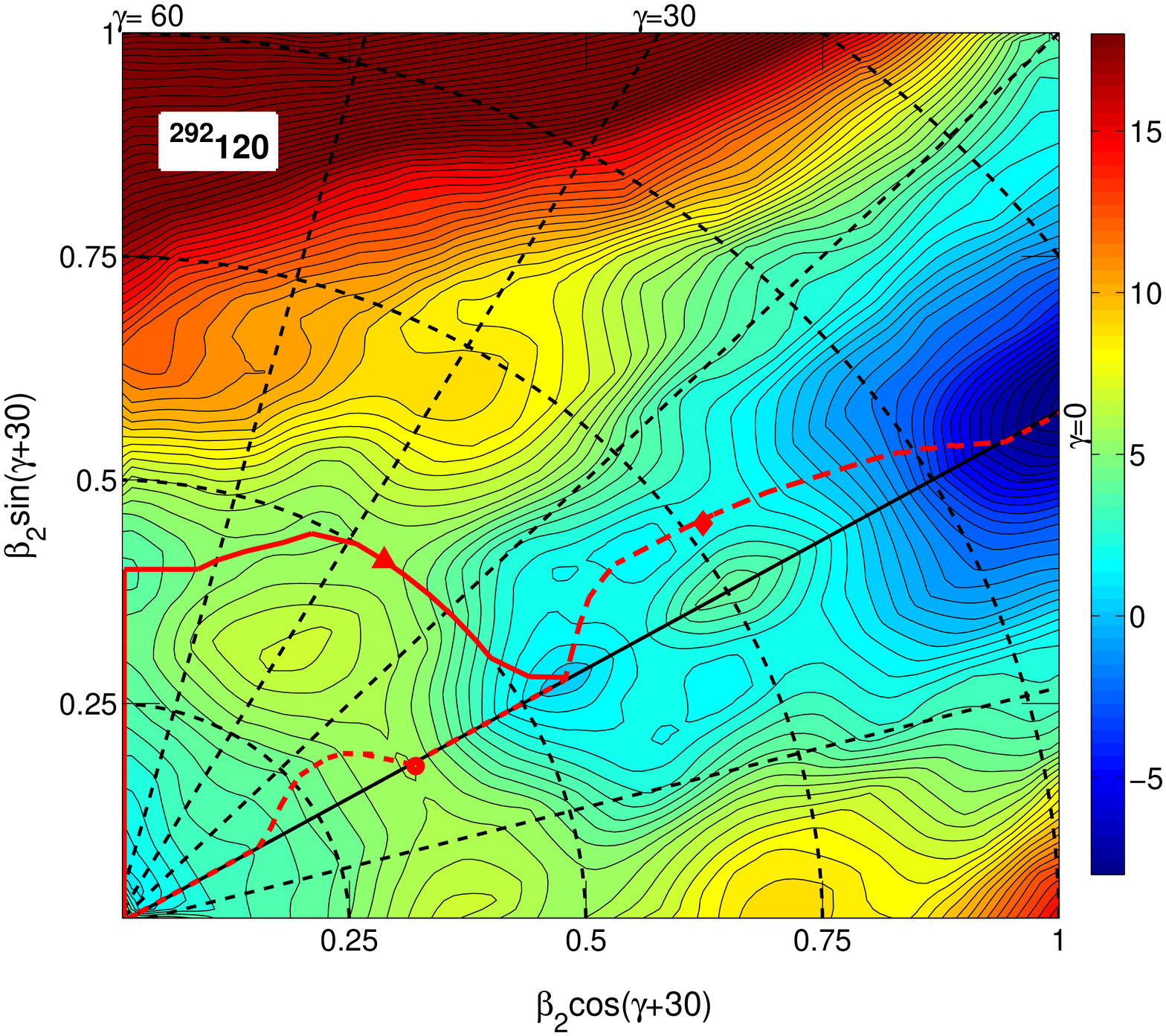}
\hspace{-0.2cm}
\includegraphics[width=8.0cm,angle=0]{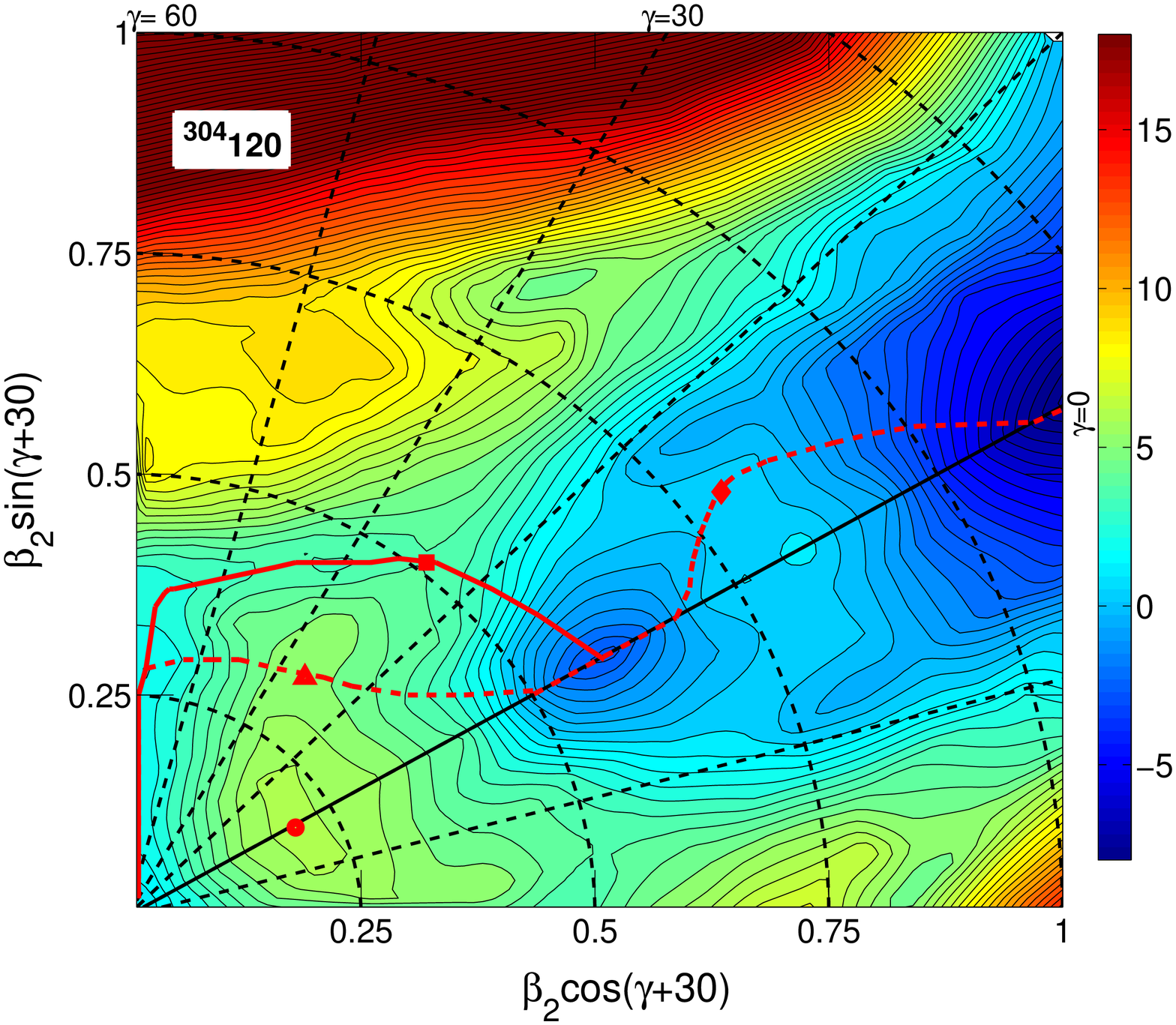}
\includegraphics[width=8.0cm,angle=0]{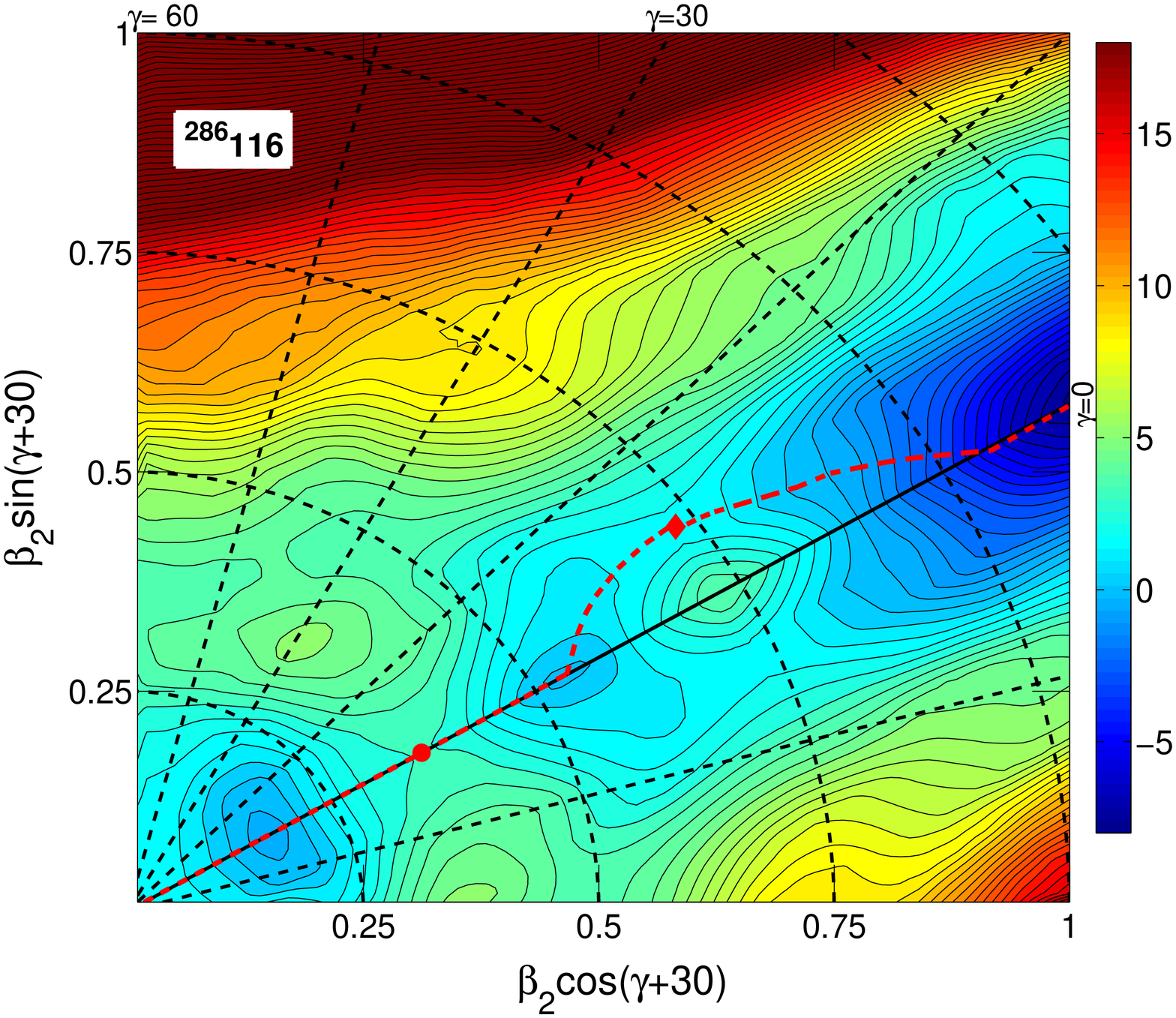}
\hspace{-0.2cm}
\includegraphics[width=8.0cm,angle=0]{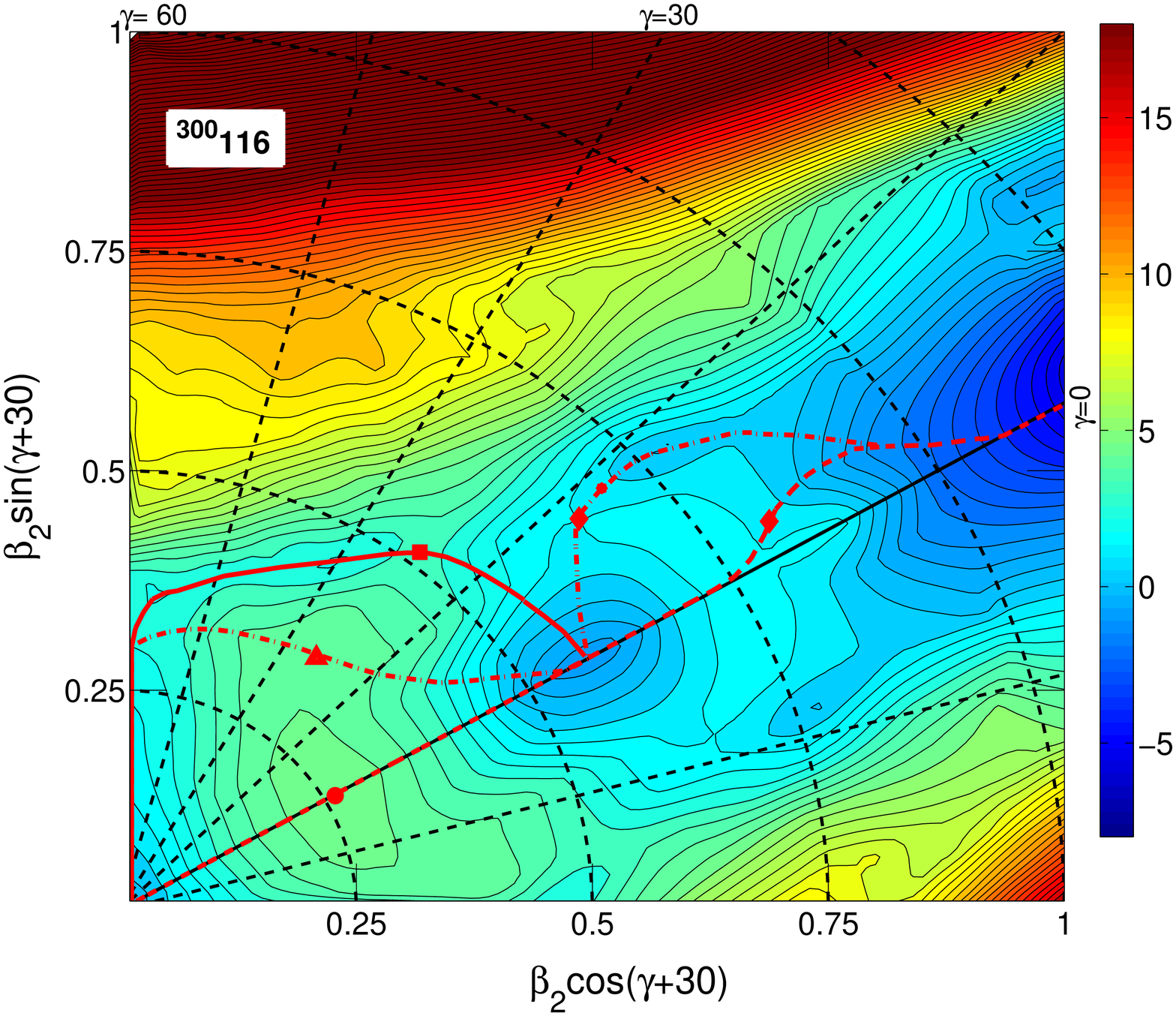}
\includegraphics[width=8.0cm,angle=0]{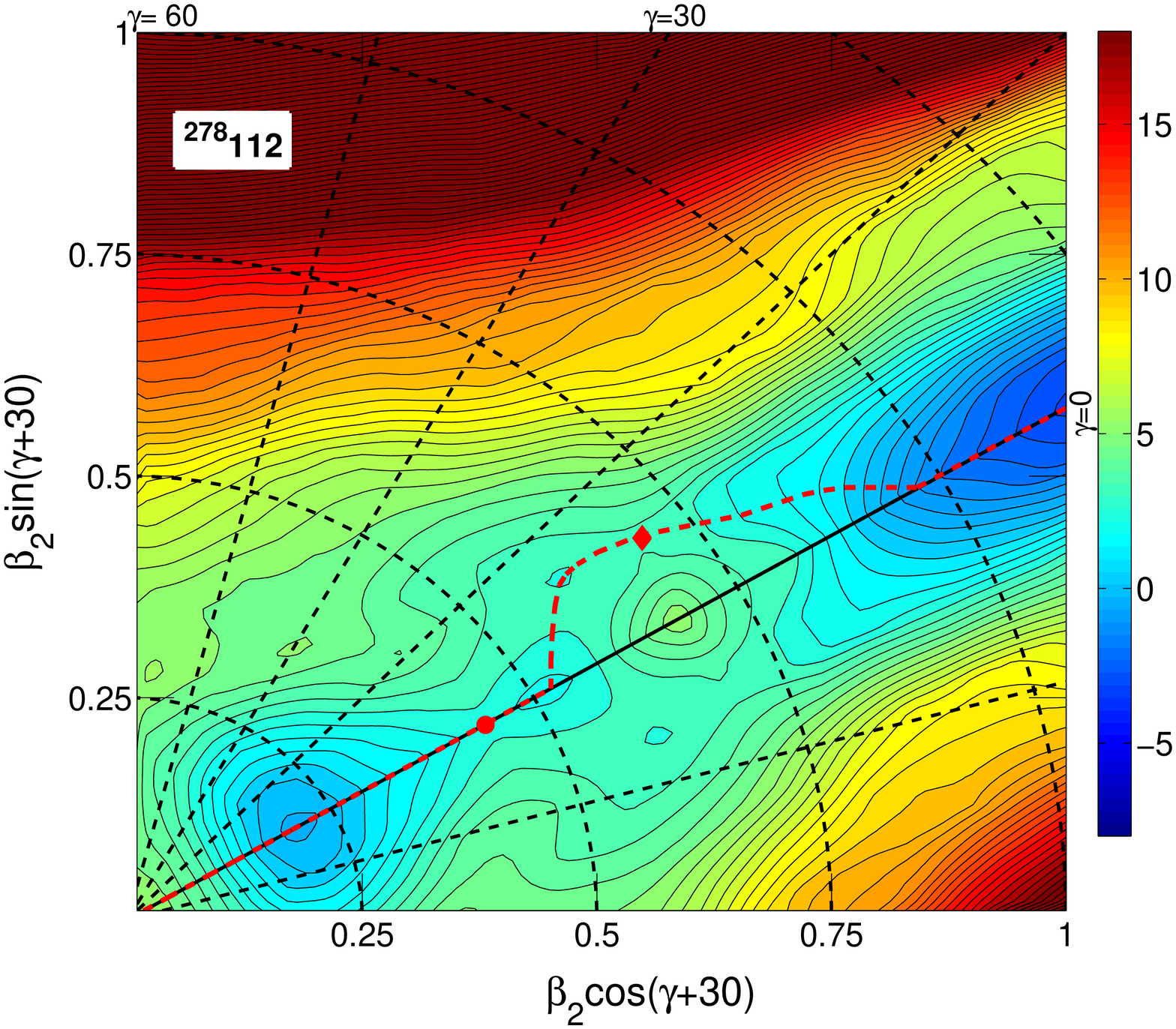}
\hspace{-0.2cm}
\includegraphics[width=8.0cm,angle=0]{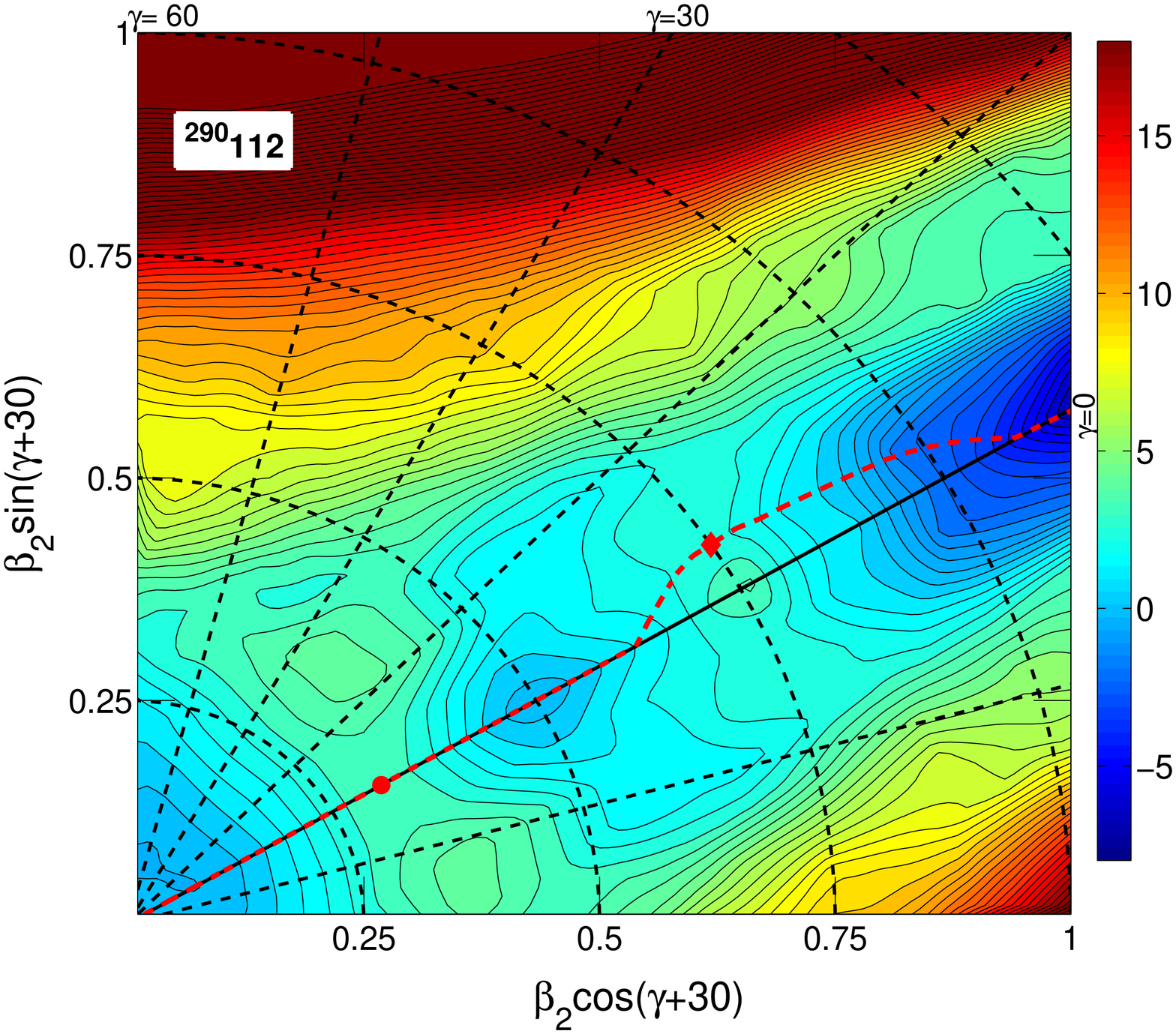}
\end{center}
\caption{Potential energy surfaces of selected nuclei. The energy
difference between two neighboring equipotential lines is equal to
0.5 MeV. The saddles along the 'Ax', 'Tr-A', and 'Tr-B' fission pathes
are shown by solid circles, triangles, and squares, respectively. The 
solid diamonds show the outer fission barrier saddles. The saddles are
defined via the immersion method \cite{MSI.09}, while the fission pathes 
as minimum energy pathes \cite{ERE.07,STH.08} which represent the most 
probable pathway connecting two minima via a given saddle. Note that contrary 
to other nuclei there are two possible fission pathes for outer fission 
barrier in the $^{300}$116 nucleus. }
\label{pes-2D-6panel}
\end{figure*}

  All mean-field calculations for fission barriers with triaxiality included
are computationally demanding. This is the reason why simple pairing interactions
(either seniority pairing with constant strength $G$ or the zero-range $\delta$-force)
and the BCS framework are used in the majority of the calculations (see Table
\ref{tab4}). The overview given in this table focuses mainly on the
investigations including triaxial deformation and on the most recent and
comprehensive studies within the specific theoretical frameworks. In particular,
the studies of fission barriers in superheavy elements (SHE) are included
irrespective of whether the calculations  include triaxial  deformation or
not. This table shows that the majority of studies neglect particle number
projection and use different prescriptions for the size of the pairing window.
The impact of these prescriptions on the height of the inner fission barrier
has recently been investigated in Ref.\ \cite{KALR.10}.

The inclusion of triaxiality has improved the accuracy of the description
of the inner fission barriers in actinides in all state-of-the-art models~\cite{DGGL.06,DPB.07,MSI.09,AAR.10}.
Figs.\ \ref{N-dep} and \ref{Z-dep} show the differences
between experimental and calculated heights of inner fission barriers obtained
in different theoretical models as a function of neutron and proton numbers,
respectively. Note that this comparison covers only results of systematic
triaxial calculations which simultaneously include even-even Th, U, Pu, Cm and
Cf nuclei. To our knowledge, no such calculations have been published with DFT
based on Skyrme forces. As a result, these figures cover all existing systematic
triaxial studies of inner fission barriers in actinides.

The $\delta$-values displayed on the panels of Figs.\ \ref{N-dep} and
\ref{Z-dep} show the average deviation from experiment for the calculated
heights of inner fission barriers. One can see that they are of the same magnitude
in the different approaches and minor differences between the approaches in the
$\delta$-values are not important considering the considerable uncertainties
in the extraction of inner fission barrier heights from experimental data
as seen, for example, in the differences of the compilations of Refs.\
\cite{RIPL-2} and \cite{MM.11}.

However, the similarity of the average trends of these deviations (shown by
thick dashed lines in Figs.\ \ref{N-dep} and \ref{Z-dep}) as a function of
neutron and proton numbers is more important considering the differences in
underlying mean fields and in the treatment of pairing correlations.
At present, it is difficult to find a clear explanation for these trends.
Although differences in the treatment of pairing correlations (BCS
with monopole pairing and of different pairing windows in the
CDFT \cite{AAR.10} and MM \cite{DPB.07,MSI.09} calculations versus the
Hartree-Fock-Bogoliubov framework based on the D1S force in
Gogny DFT \cite{DGGL.06}) can contribute to deviations between theory
and experiment \cite{KALR.10}, it is quite unlikely that they are responsible
for the observed trends of the deviations.

\section{Results and Discussion}
\label{res+diss}

  The nucleus $^{292}$120 is predicted to be a spherical doubly magic nucleus
in CDFT \cite{BRRMG.98,AKF.03}. Its potential energy surface in the
$\beta-\gamma$ plane is shown in Fig.\ \ref{pes-2D-6panel}. It is interesting to
compare it with the PES of the nucleus $^{240}$Pu shown in Fig.\
\ref{pes-2D}. These two PES's are representative examples of typical
PES's in actinides and superheavy nuclei. The gross structure of these two
PES's is defined by the fact that the total energy is generally increasing
when moving away from the $\gamma=0^{\circ}$ axis; so it looks like a canyon.
However, there are local structures inside the canyon which define the
differences between the two mass regions with respect to the impact of
triaxiality on the inner and outer fission barriers.

  In $^{240}$Pu, a large hill is located at  the axial shape $\beta_2 \sim 0.5$
inside a canyon. As a consequence, the fission path from  the normal deformed
minimum initially proceeds along the $\gamma =0^{\circ}$ axis, then bypasses the axial
$\beta_2$=0.5 hill via a path with $\gamma \sim 10^{\circ}$, and then proceeds
along the bottom of the canyon on an axially symmetric path again. As a result of this
bypass, the inner fission barrier heights of  the actinides are lowered by $1-4$
MeV due to triaxiality \cite{AAR.10}. However, the calculated outer fission
barriers of the actinides are not affected by triaxiality \cite{AAR.10}.

\begin{figure}
\includegraphics[angle=0,width=8.0cm]{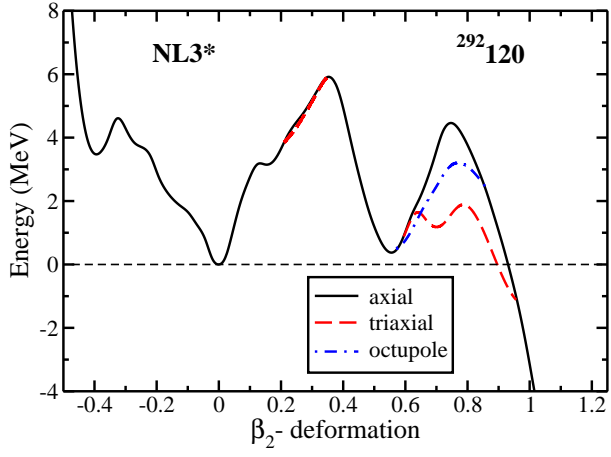}
\caption{ (Color online) Deformation energy curves for the $Z=120,\,N=172$
nucleus obtained with the NL3* parametrization. The black solid, red
dashed and blue dot-dashed lines display the deformation energy curves for
the axially symmetric, triaxial and axial octupole deformed solutions.
We show the deformation energy curves for the last two solutions
only in the range of $\beta_2$ values
where it is lower in energy than the deformation energy curve of the
axially symmetric solution. Note that the 'Tr-B' fission path via
saddle at $\beta_2=0.47, \gamma=26^{\circ}$ (see text for details)
is not shown since this saddle is lower than the one of the 'Ax' fission
path only by 0.2 MeV (see Table \ref{tab5}); the results of Ref.\
\cite{GSPS.99} suggests that spontaneous fission exploiting this path is
less likely than the one along the 'Ax' fission path.}
\label{pot}
\end{figure}

The properties of the PES of the nucleus $^{292}$120, defining the fission
pathes, are completely opposite to the case of $^{240}$Pu since there are two triaxial
and one axial hills inside the PES canyon of $^{292}$120. Two triaxial hills
are located at moderate deformations ($\beta_2 \sim 0.35,\,\,\gamma \sim
\pm30^{\circ}$), while the axial hill is superdeformed ($\beta_2 \sim 0.75$).
The fission path (shown by red dashed line in Fig.\
\ref{pes-2D-6panel}) starts at a spherical shape, then proceeds between
two triaxial hills ($\beta_2 \sim 0.35,\,\,\gamma\sim $30$^{\circ}$) and
bypasses the axial hill at $\beta_2\sim$0.75 via a $\gamma \sim 7^{\circ}$
path. The $\gamma$-softness of the PES, which exists between the two triaxial
hills, has only a minor effect on the height of the inner fission barrier;
the triaxial solution is lower than the axial one by 100-200 keV at
$\beta_2=0.2-0.3$ deformations (see Fig.\ \ref{pot}). However, this figure
shows that {\it along this fission path} the height of inner fission barrier
is not affected by triaxiality.

The second fission path shown by a solid red line starts at spherical shape,
and proceeds along the axially symmetric $\gamma=60^{\circ}$ axis, via a
saddle point at ($\beta_2 \sim 0.47$, $\gamma \sim 21^{\circ}$) and
then along the first fission path after second minimum. The existence
of the valley between the walls of the canyon and triaxial hills at
($\beta_2 \sim 0.35,\,\,\gamma \sim \pm30^{\circ}$) is a necessary condition
for the existence of this path. This type of fission path exists in a number
of nuclei, so for convenience we will call it as triaxial 'Tr-B' fission
path. The unusual physical feature of this second path is the fact that
initially the nucleus has to be squeezed along the axis
of symmetry, thus creating an oblate nucleus with quadrupole deformation
$\beta_2 \sim -0.39$. This is contrary to the usual picture of fission where
the prolate or near-prolate nucleus is stretched out along the path of
increasing quadrupole deformation. Note that the saddle of this fission path
is only by 0.2 MeV lower than the saddle of the first fission path (see
Table \ref{tab5}) and in addition it is much longer. Therefore, considering
a dynamical calculation taking into account the action integral along the
entire fission path~\cite{GSPS.99} this path will probably not
contribute much to the fission probability.

  Contrary to the actinides, the triaxiality has a considerable impact
on the shape and the height of outer fission barrier which is lowered by
$\sim$ 3 MeV in $^{292}$120 nucleus (see Fig.\ \ref{pot}). {Note, that
in this nucleus,} the lowering of outer fission barrier due to octupole 
deformation is substantially smaller than the one due to triaxiality.

It is also evident that the landscape of the PES and the existence of
saddle points and valleys depends on the proton and neutron numbers.
This is clearly seen in Fig.\ \ref{pes-2D-6panel} and in Table \ref{tab5}.
In the $Z=120$ isotopes, the increase of neutron number up to $N=184$
and beyond
leads to the emergence of an axial hill at $\beta_2 \sim 0.2-25$; this is
clearly visible in the nucleus $^{304}$120 in Fig.\ \ref{pes-2D-6panel}.
Its existence in high-$Z$ and neutron-rich nuclei leads to the shift
of the first fission path in the deformation space. In lighter systems
this path proceeds via the axially symmetric saddle as, for instance, in
the nucleus $^{292}$120 in Fig.\ \ref{pes-2D-6panel}. For heavier
systems it starts at spherical shape\footnote{As discussed later,
some of the nuclei have superdeformed ground states. However, 
the calculated outer fission barriers are rather small ($\sim 2$ 
MeV relatively to the superdeformed minimum) which suggests that these 
states are extremely unstable against fission. Thus, we start the discussion 
from spherical/weakly(normal)-deformed minima which have better survival 
probability against fission.}, proceeds along the axially
symmetric $\gamma=60^{\circ}$ axis up to $\beta_2 \sim 0.25$ and then via
a saddle point at ($\beta_2 \sim 0.32, \gamma \sim 26^{\circ}$), located
between a triaxial 
($\beta_2 \sim 0.35,\,\,\gamma\sim 28^{\circ}$)\footnote{This
hill is clearly visible when the energy difference between two 
neighboring equipotential lines is set to 0.1 MeV, see Fig.\ \ref{302-valley} 
below.}
and an axial $\beta_2 \sim 0.2$ hill, to the axially symmetric superdeformed
minimum. This type of fission path exists in a number of nuclei. We
will label it as triaxial 'Tr-A' for convenience. In addition, there is a
second triaxial path 'Tr-B' via a saddle at $\beta_2\sim 0.5, \gamma
\sim 22^{\circ}$, which is similar to the one present in the nucleus
$^{292}$120.

  The structure of the PES of the nucleus $^{300}$116 is similar to the
one of the nucleus $^{304}$120 in Fig.\ \ref{pes-2D-6panel}. The lowering
of neutron number below $N=184$ in the nuclei under study leads to the
situation that the fission path via the axially symmetric inner saddle
(called as 'Ax' below) becomes energetically favored.  This is clearly
seen in the PES of the $^{286}$116, $^{278}$112 and $^{290}$112 nuclei in
Fig.\ \ref{pes-2D-6panel} and in Table \ref{tab5}.

  It is clear that the landscape of the PES of superheavy nuclei
in the region of the inner fission barrier is more complicated than in
the case of the actinides. The energetically favored fission path
in the majority of these nuclei proceeds through an axially symmetric
saddle (see Fig.\ \ref{pes-2D-6panel} and Table \ref{tab5}). Only
in a few nuclei ($^{304,306,308}$120, $^{300,302,304}118$ and
$^{300}$116), triaxial fission path 'Tr-A' is lower in energy than
axial 'Ax' fission path. The triaxial fission path 'Tr-B' exists in
the majority of the nuclei apart from neutron-poor $Z=114$ nuclei and
$Z=112$ nuclei (see Fig.\ \ref{pes-2D-6panel} and Table \ref{tab5}).
The energy of the saddle point of this fission path is lower than
the one of the first fission path (either 'Ax' or 'Tr-A') in a number
of nuclei. However, the results of calculations within the MM
method \cite{GSPS.99} suggest that although triaxiality lowers the static
fission barriers, it plays a minor role in spontaneous fission of
superheavy nuclei with $Z\leq 120$. This is because a fission path
via an oblate shape and triaxial saddles is substantially longer as
compared to the axially symmetric path which leads to a significant
reduction of the penetration probability. However, this complexity of
the PES and the presence of two fission pathes for the inner fission
barriers calls for finding the dynamical path along which the fission
process takes place in CDFT. Work in this direction is in progress.

\begin{figure*}
\includegraphics[angle=0,width=16.0cm]{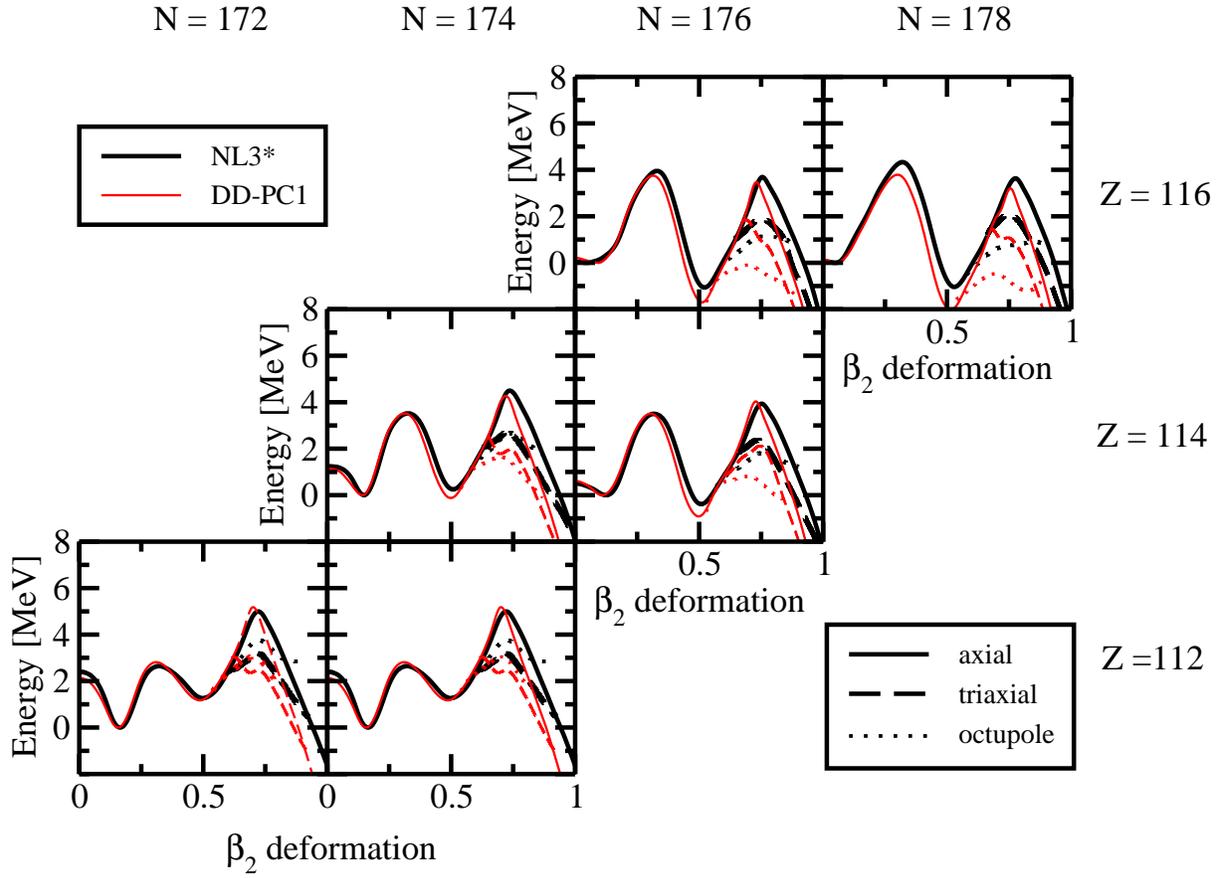}
\caption{(Color online) Deformation energy curves for the $Z=112,\,114$ and
116 nuclei obtained with the NL3* and DD-PC1 parameterizations of the RMF
Lagrangian. Solid lines correspond to axial solutions with reflection symmetry 
(A), dashed lines to triaxial solutions with reflection symmetry (T), and 
dotted lines to octupole deformed solutions with axial symmetry (O). Note that 
the T and O solutions are shown only in the deformation range in which they 
are lower in energy than axial solution.}
\label{pot3a}
\end{figure*}

\begin{figure*}
\includegraphics[angle=0,width=16.0cm]{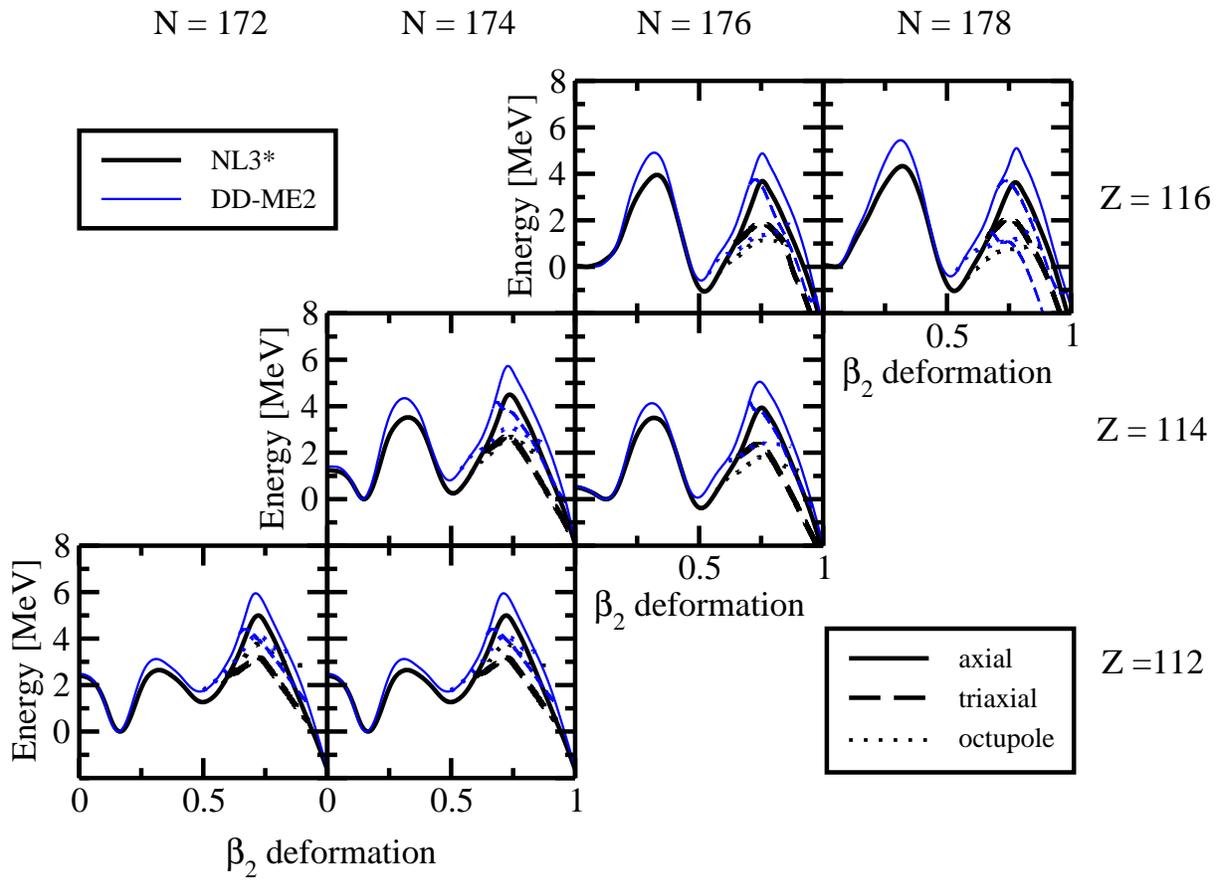}
\caption{ (Color online) The same as Fig.\ \ref{pot3a} but for the NL3*
and DD-ME2 parametrizations.}
\label{pot3b}
\end{figure*}

\begin{figure*}[ht]
\begin{center}
\includegraphics[width=8.0cm,angle=0]{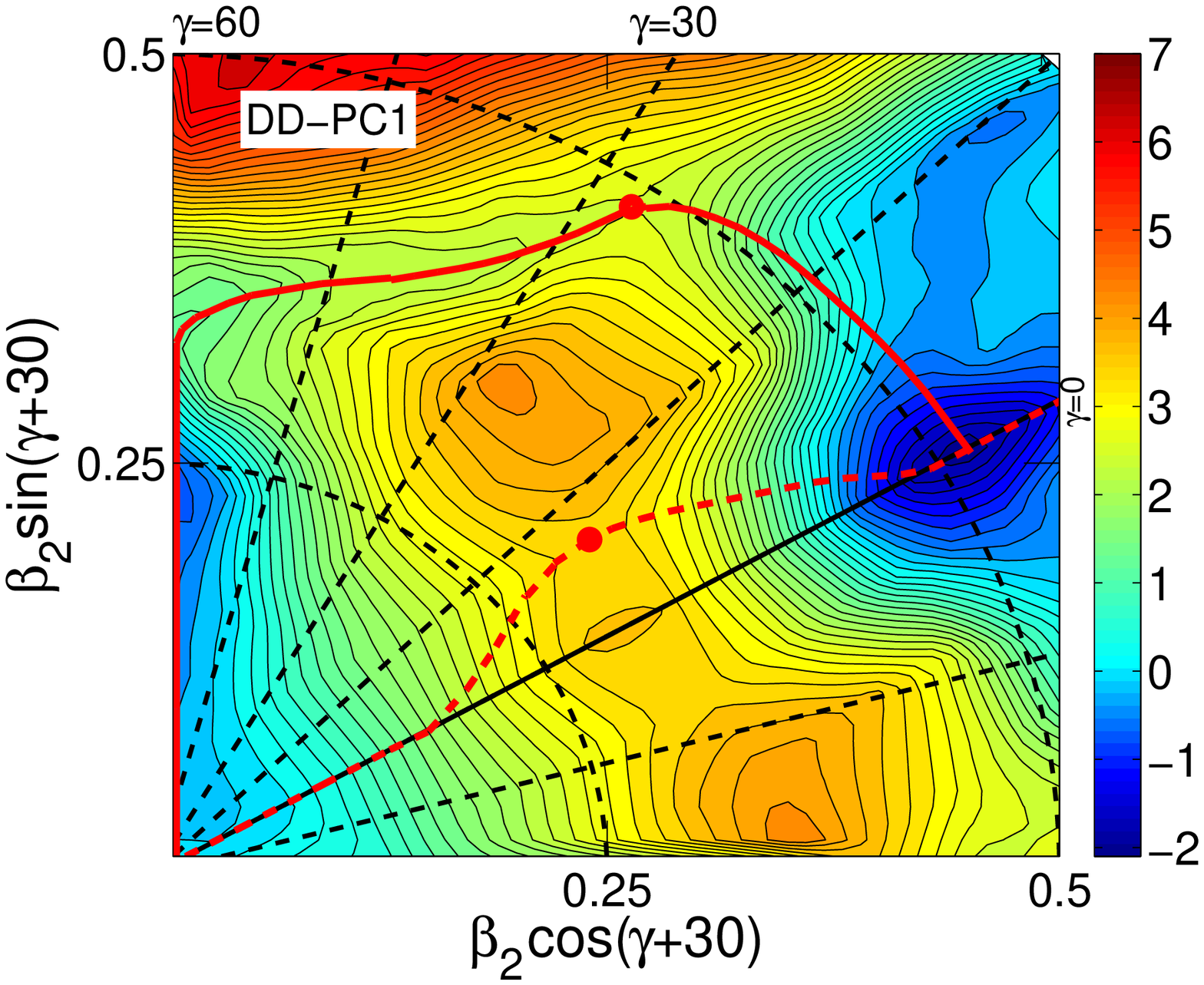}
\hspace{-0.2cm}
\includegraphics[width=8.0cm,angle=0]{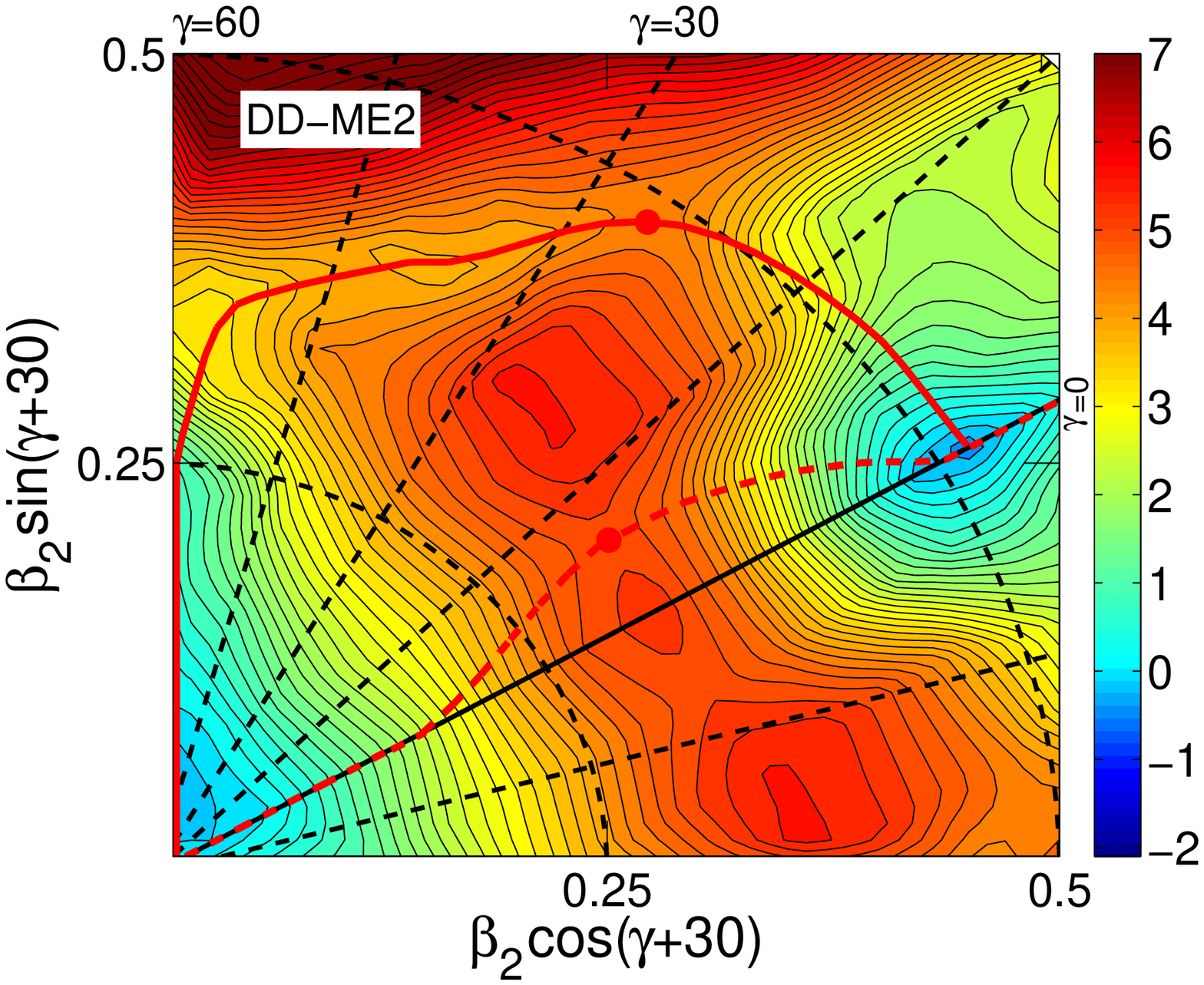}
\includegraphics[width=6.3cm,angle=-90]{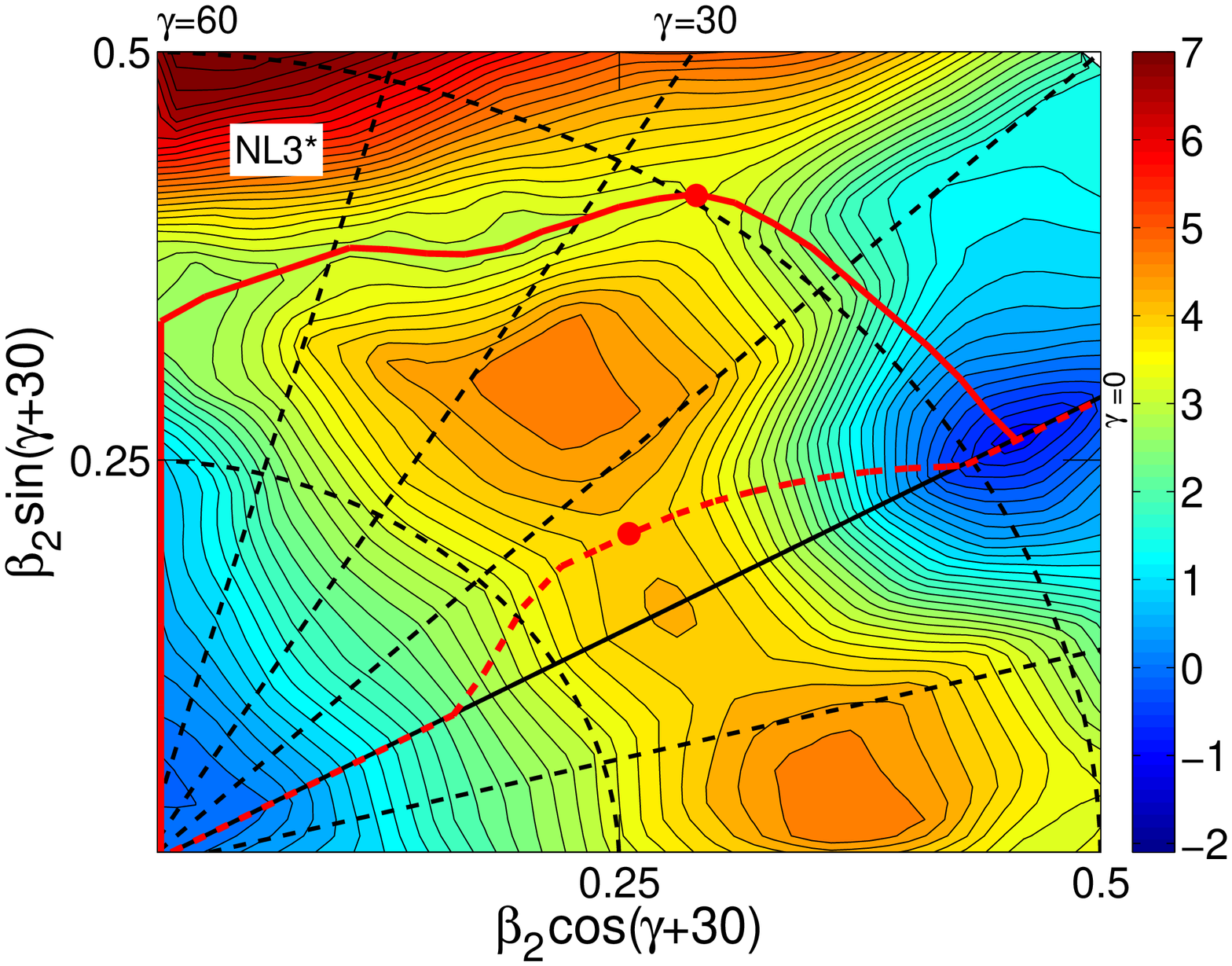}
\end{center}
\caption{The same as in Fig.\ \ref{pes-2D-6panel} but for the nucleus 
$^{294}$116. The results are shown for the parametrizations NL3*, DD-PC1
and DD-ME2. The energy difference between two neighboring
equipotential lines is equal to 0.2 MeV.}
\label{pes-2D-dif-param}
\end{figure*}

\begin{figure}[ht]
\includegraphics[angle=0,width=8.0cm]{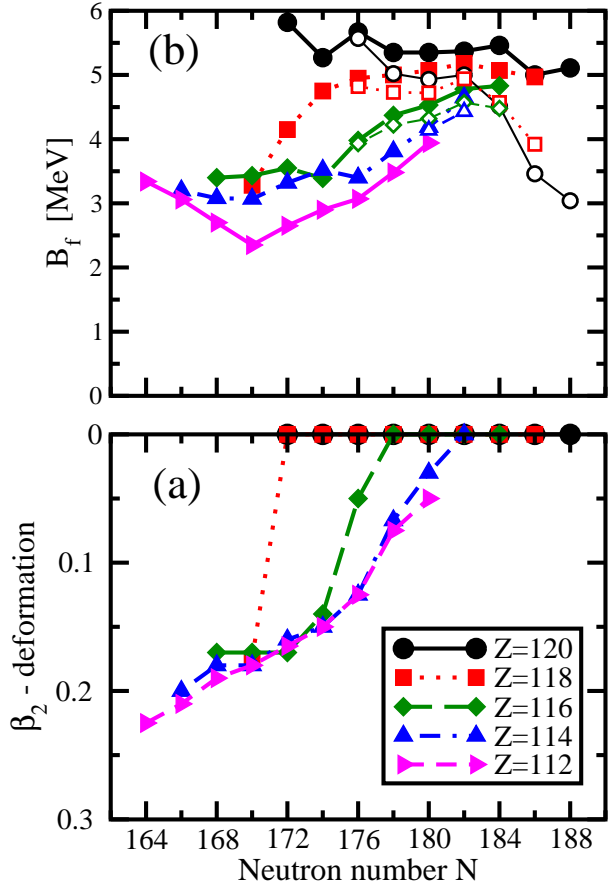}
\caption{(Color online) 
(panel (a)) The deformations of the ground states of even-even 
$Z=112-120$ nuclei as a function of neutron number.
(panel (b)) The energies of axial (large solid 
symbols, thick lines) and triaxial (either 'Ax-Tr' or 'Tr-A'; shown by 
small open symbols and thin lines) saddles with respect to the 
spherical/weakly(normal)-deformed minima  in  these nuclei 
as a function of neutron number. Note that the same type of symbols is
used for axial and triaxial saddles in a given isotope chain.}
\label{pot2}
\end{figure}

  Figs.\ \ref{pot3a} and \ref{pot3b} show deformation energy curves for several
isotopes with charge numbers $Z=112,\,114$ and 116 for three classes of  CDFT
models. Experimental estimates of the inner fission barrier heights were obtained for
these nuclei in Ref.\ \cite{IOZ.02}. The potential energy structure of these
nuclei is similar to the one seen in $^{292}$120 (Fig.\ \ref{pes-2D-6panel});
the presence of two triaxial hills at moderate deformations ($\beta_2 \sim 0.35,\,\,\gamma
\sim \pm30^{\circ}$) defines two possible fission pathes: one between the
hills (either 'Ax' or 'Ax-Tr' pathes) and another between the hill and the
walls of the PES canyon ('Tr-B' path). The $\gamma$-softness of the PES, which
exists between the two triaxial hills, defines whether the fission path passing
between these two hills is fully axially symmetric ('Ax' path) or has some
degree of triaxiality ('Ax-Tr' path). The 'Ax-Tr' path is generally similar
to the 'Ax' path but differs from it by the fact that moderate triaxiality
$(\gamma \leq 10^{\circ})$ appears at the saddle and along the shoulder of
inner fission barrier. Although the heights of the discussed hills in
the region of inner fission barrier depends on the parametrization, the
observed features of the potential energy surfaces and fission pathes are rather
independent of the parametrization (Fig.\ \ref{pes-2D-dif-param}).
Tables \ref{tab5} and \ref{tab7} show that the saddle
point along this path has triaxiality only in the case of the two
nuclei $^{292,294}$114. However, even in these cases the 'Ax' saddle is
only 50-150 keV higher than the 'Tr-A' saddle. Thus, similarly to the
$^{292}$120 nucleus triaxiality only marginally affects the inner fission
barriers. Note that the ground states are somewhat deformed in these nuclei
(see Figs.\ \ref{pot3a} and \ref{pot2}a).

  For these nuclei triaxiality has a considerable impact on the shape
and height of the outer fission barriers; the decrease of the heights
of the outer  fission barriers due to triaxiality is typically in the
range of 1.5-2.0 MeV and it depends on the particle number and on the
RMF parametrization.  Note that the outer fission barriers are affected 
also by octupole deformation. In the $Z=112$ nuclei, the triaxial saddle 
is lower in energy than the octupole saddle. The situation is reversed 
in the $Z=114,116$ nuclei.

  Among the different classes of CDFT models, the DD-ME2 parametrization
always gives the highest values for the inner and outer fission barrier
heights. They are (on average) by 1 and 1.5 MeV higher than the ones
obtained in the NL3* and the DD-PC1 parametrizations. The heights 
and the shapes of the inner fission barriers are very similar for the 
NL3* and DD-PC1 parametrizations. The outer fission barriers also come 
close to each other in these two parametrizations in axial [A], triaxial [T] 
and octupole [O] (for $Z=112$ nuclei) calculations.  However, with 
increasing of $Z$ from 114 to 116 the difference between the results
of octupole calculations increases; the difference between the energies
of the octupole saddles obtained in the NL3* and DD-PC1 parametrizations
reaches 2 MeV in the $Z=116$ nuclei (see Fig.\ 
\ref{pot3a}).

  There is only one experimental work \cite{IOZ.02} where
estimates on the heights of inner fission barriers in superheavy nuclei with
$Z=$ 112, 114, and 116 have been obtained. Unfortunately, experimentally the fission
barriers are accessible only indirectly and a model-dependent
analysis is used to obtain these quantities, which causes an
ambiguity in the comparison with theoretical results. Even in the actinide region
where the fission barrier heights were extracted from a number
of independent experiments with high statistics (see, for example, Ref.\ \cite{MSI.09}),
a typical uncertainty in the experimental values, as suggested by the differences among various
compilations, is of the order of $\pm0.5$ MeV \cite{SGP.05}. These uncertainties are expected to
be higher in superheavy nuclei since the estimates of Ref.\ \cite{IOZ.02} are based on
experimental data represented by low statistics and on a method
which differs from the methods used in the analysis of fission barrier heights in actinides.
In addition, there is no independent confirmation of the inner fission barrier height
estimates of Ref.\ \cite{IOZ.02}. The interpretation of experimental data
based on cross sections in terms of fission barrier heights becomes even more
complicated when the fission path has a double hump structure,
which according to many calculations may be the case in superheavy
nuclei. The widening of the barrier due to the second hump
(or its remnant) would require the lowering of the inner fission barrier height;
this possibility has not been taken into account in the analysis of Ref.\ \cite{IOZ.02}.
Based on this discussion it is clear that the level of confidence of fission
barrier height estimates for superheavy nuclei is significantly lower than the one
for the actinides.

  According to Ref.\ \cite{IOZ.02}, the estimated lower limits for fission barrier
heights in even-even $Z=112, 114$ and 116 nuclei shown in Fig.\ \ref{pot3a} and
\ref{pot3b} are
5.5, 6.7 and 6.4 MeV, respectively. Our results for the heights of inner fission
barrier in these nuclei along the 'Ax' and 'Ax-Tr' fission pathes are always
smaller than the experimental data by $1-3$ MeV. The saddle along the 'Tr-B'
fission path is somewhat lower in energy (by 0.4-0.7 MeV) than the saddle along
the 'Ax-Tr' path in $^{292,294}$116 nuclei (see Table \ref{tab7}). However, as
discussed above (based on the results of Ref.\ \cite{GSPS.99}) it is
unlikely that the fission predominantly proceeds along the 'Tr-B' path
considering especially (i) that the gain in saddle point energy as compared
with the  'Ax-Tr' path is small and (ii) the 'Tr-B' path is substantially
longer as compared with the 'Ax-Tr' path.

Considering the discussion above, it is not clear at this moment how serious
the discrepancy {\it between calculations and experimental estimates} is. However,
the results of the calculations suggest one possible way to increase the heights
of inner fission barriers. Potential energy surfaces in the ground state region
of these nuclei are extremely soft (see Figs.\ \ref{pot3a} and \ref{pot3b}). For
such nuclei, the correlations beyond mean field taken into account, for example,
by the generator coordinate method can lower the energy of the ground state by a
few MeV without affecting much the barrier top, thus effectively increasing the
height of inner fission barrier.

   Systematic calculations of inner fission barriers have been performed for
even-even $Z=112-120$ nuclei with $N-Z=52-68$ using the NL3* parametrization.
This is the same parametrization which has been used in systematic calculations
of fission barriers in actinides~\cite{AAR.10} and which provides 0.76 MeV
average deviation from experiment for the heights of inner fission barriers in these
nuclei. For a given proton number $Z$ a sequence of nine even-even nuclei is
selected such that the middle nucleus of this sequence roughly
corresponds to an experimentally observed nucleus (or a nucleus
which is expected to be observed in near future). This crudely
outlines the region which may be experimentally studied within
the next one or two decades. The results of these calculations are
summarized in Table \ref{tab5}.

 Based on the results of Ref.\ \cite{GSPS.99}, it is reasonable to
neglect the path 'Tr-B' since it is substantially longer as compared with
other pathes which leads to significant reduction of the penetration
probability. In Fig.\ \ref{pot2}b we consider the evolution of the heights
of the inner fission barriers as a function of the neutron number $N$.
The $Z=112,\, 114$ and 116 isotope chains show a generally increasing 
trend for the barrier heights with increasing neutron number for
$N\geq 172$. Fig.\ \ref{pot2}a suggests that the origin of this trend can be 
traced back to the deformation
of the ground state. For small neutron numbers these nuclei are deformed in
the ground state. However, they gradually become spherical when approaching
$N=184$ because there is a spherical shell gap at this neutron number
(see, for example, Fig.\ 28 in Ref.\ \cite{AKF.03}). The negative
shell correction energy at the ground state is larger in absolute value
in the vicinity of the $N=184$
spherical shell gap than at lower neutron numbers, where the ground state
is deformed and characterized by a larger level density in the vicinity
of the Fermi level. The level density (and, as a consequence, the shell
correction energy) at the saddle point of the inner fission barrier does
not change so drastically as the one at the ground state. As a result, the
heights of inner fission barriers, which are defined as the energy differences
between the binding energies of the ground state and saddle point, show
the observed features. Note that in these nuclei the energies of the
'Ax' and 'Ax-Tr' saddles differ by at most 200 keV, the difference between
the energies of the 'Ax' and 'Tr-A' saddles does not exceed 400 keV.

 The nuclei with $Z=$ 118 and 120 have, with the exception of the nucleus $^{288}118$,
spherical minima due to the presence of the $Z=120$ spherical shell
gap. The nucleus $^{292}120$ has the highest value for the fission barrier among
all nuclei under investigation. This is connected with its doubly magic nature
in CDFT. In the isotope chain with $Z=120$, moving away from the $N=172$ shell closure,
shell effects connected with spherical shape become less pronounced and this leads
to a decrease of the inner fission barrier, because the barrier height in these
nuclei is defined with respect to the spherical ground state. Apart from the lightest
two and the heaviest $Z=118$ isotopes, the fission barrier heights of the $Z=118$ chain
are nearly constant as a function of neutron number and they are close to 4.5 MeV.
Note that in these nuclei the energies of the 'Ax' and 'Ax-Tr' saddles differ only by
100-400 keV. However, the difference between the energies of the 'Ax' and 'Tr-A' saddles
can reach 2 MeV (as in the case of the nucleus $^{308}$120, see Table \ref{tab5}).

It is important to mention that the valley between the axial $\beta_2\sim 0.2$ and
triaxial $\beta_2\sim 0.40, \gamma\sim 25^{\circ}$ hills leading to a 'Tr-A' saddle
is rather shallow. Its depth with respect to the tip of the triaxial hill varies
between 100 and 300 keV. The latter value is obtained, for example, in $^{302}$120
nucleus, see Fig.\ \ref{302-valley}.

\begin{table}[ptb]
\caption{The heights of axially symmetric ('Ax') and triaxial ('Ax-Tr',
'Tr-A' and 'Tr-B') saddle points [in MeV] with respect to
spherical/weakly(normal)-deformed minima
and their deformations. Note that only quadrupole
deformation is given for axial saddle. Columns 4 and 5 show the values for the
saddles of either 'Ax-Tr' or 'Tr-A' fission pathes. The asterisk to the fission
barrier height in column 4 indicates that the values for the 'Ax-Tr' path are
displayed; the absence of the asterisk implies that the values for the 'Tr-A' path
are shown. Note that triaxial saddles are shown only in the cases when their heights
are lower than those of axial saddle.
\label{tab5}}
\begin{center}
\renewcommand{\arraystretch}{1.2} 
\begin{tabular}{|c|c|c|c|c|c|c|}
\hline
Nucleus & $B_f^{Ax}$ & $\beta_2$ & $B_f^X$ & ($\beta_2$,$\gamma$) & $B_f^{Tr-B}$ &
($\beta_2$,$\gamma$)\\ \hline\hline
1           & 2  & 3  & 4  & 5 & 6 & 7 \\ \hline
\multicolumn{7}{|c|}{\bf $Z=120$ nuclei} \\ \hline
$^{308}$120 & 5.11 & 0.20 & 3.04 & (0.34,26$^{\circ}$) & 2.23 & (0.48,22$^{\circ}$)\\
$^{306}$120 & 5.00 & 0.20 & 3.46 & (0.33,26$^{\circ}$) & 2.02 & (0.50,22$^{\circ}$)\\
$^{304}$120 & 5.46 & 0.21 & 4.51 & (0.33,26$^{\circ}$) & 3.30 & (0.51,21$^{\circ}$)\\
$^{302}$120 & 5.37 & 0.19 & 5.00 & (0.33,25$^{\circ}$) & 3.93 & (0.50,24$^{\circ}$)\\
$^{300}$120 & 5.35 & 0.32 & 4.93*& (0.31,11$^{\circ}$) & 4.08 & (0.50,26$^{\circ}$)\\
$^{298}$120 & 5.35 & 0.33 & 5.02*& (0.33,12$^{\circ}$) & 4.27 & (0.49,28$^{\circ}$)\\
$^{296}$120 & 5.67 & 0.34 & 5.57*& (0.35,10$^{\circ}$) & 4.92 & (0.50,27$^{\circ}$)\\
$^{294}$120 & 5.27 & 0.34 &      &                     & 4.73 & (0.50,27$^{\circ}$)\\
$^{292}$120 & 5.82 & 0.36 &      &                     & 5.61 & (0.50,27$^{\circ}$)\\ \hline \hline
\multicolumn{7}{|c|}{\bf $Z=118$ nuclei} \\ \hline
$^{304}$118 & 4.97 & 0.21 & 3.92 & (0.33,26$^{\circ}$) & 3.03 & (0.51,26$^{\circ}$)\\
$^{302}$118 & 5.07 & 0.23 & 4.57 & (0.33,26$^{\circ}$) & 3.41 & (0.51,25$^{\circ}$)\\
$^{300}$118 & 5.18 & 0.28 & 4.94*& (0.27,15$^{\circ}$) & 3.37 & (0.51,26$^{\circ}$)\\
$^{298}$118 & 5.07 & 0.32 & 4.72*& (0.32,12$^{\circ}$) & 3.67 & (0.50,27$^{\circ}$)\\
$^{296}$118 & 5.00 & 0.32 & 4.73*& (0.33,10$^{\circ}$) & 3.87 & (0.50,26$^{\circ}$)\\
$^{294}$118 & 4.95 & 0.32 & 4.82*& (0.33,10$^{\circ}$) & 4.12 & (0.50,25$^{\circ}$)\\
$^{292}$118 & 4.75 & 0.34 &      &                     & 4.35 & (0.50,23$^{\circ}$) \\
$^{290}$118 & 4.15 & 0.36 &      &                     & 4.08 & (0.49,24$^{\circ}$) \\
$^{288}$118 & 3.28 & 0.36 &      &                     &      &                     \\ \hline \hline
\multicolumn{7}{|c|}{\bf $Z=116$ nuclei} \\ \hline
$^{300}$116 & 4.83 & 0.26 & 4.48 & (0.35,25$^{\circ}$) & 3.13 & (0.52,22$^{\circ}$) \\
$^{298}$116 & 4.78 & 0.28 & 4.57 & (0.30,15$^{\circ}$) & 3.29 & (0.51,23$^{\circ}$) \\
$^{296}$116 & 4.53 & 0.32 & 4.32*& (0.32,11$^{\circ}$) & 3.35 & (0.51,24$^{\circ}$) \\
$^{294}$116 & 4.37 & 0.32 & 4.22*& (0.33,7$^{\circ}$)  & 3.40 & (0.50,25$^{\circ}$) \\
$^{292}$116 & 3.98 & 0.33 & 3.93*& (0.34,7$^{\circ}$)  & 3.39 & (0.48,26$^{\circ}$) \\
$^{290}$116 & 3.39 & 0.34 &      &                     & 3.47 & (0.48,27$^{\circ}$) \\
$^{288}$116 & 3.55 & 0.34 &      &                     &      &                     \\
$^{286}$116 & 3.43 & 0.36 &      &                     &      &                     \\
$^{284}$116 & 3.40 & 0.37 &      &                     &      &                     \\ \hline\hline
\multicolumn{7}{|c|}{\bf $Z=114$ nuclei} \\ \hline
$^{296}$114 & 4.65 & 0.28 & 4.43*& (0.31,15$^{\circ}$) & 3.42 & (0.49,24$^{\circ}$) \\
$^{294}$114 & 4.19 & 0.29 & 4.14*& (0.31,10$^{\circ}$) & 3.08 & (0.49,25$^{\circ}$) \\
$^{292}$114 & 3.81 & 0.32 &      &                     & 3.18 & (0.48,25$^{\circ}$) \\

$^{290}$114 & 3.40 & 0.34 &      &                     & 3.38 & (0.48,26$^{\circ}$) \\
$^{288}$114 & 3.52 & 0.32 &      &                     &      &                      \\

$^{286}$114 & 3.32 & 0.33 &      &                     &      &                      \\
$^{284}$114 & 3.07 & 0.38 &      &                     &      &                      \\
$^{282}$114 & 3.08 & 0.42 &      &                     &      &                      \\
$^{280}$114 & 3.20 & 0.42 &      &                     &      &                      \\ \hline
\end{tabular}
\end{center}
\end{table}

\begin{table}[ptb]
\caption{The same as Table \ref{tab5} but for $Z=112$ nuclei.}
\begin{center}
\renewcommand{\arraystretch}{1.2} 
\begin{tabular}{|c|c|c|c|c|c|c|}
\hline
Nucleus & $B_f^{Ax}$ & $\beta_2$ & $B_f^X$ & ($\beta_2$,$\gamma$) & $B_f^{Tr-B}$ & ($\beta_2$,$\gamma$)\\ \hline\hline
1           & 2  & 3  & 4  & 5 & 6 & 7 \\ \hline
\multicolumn{7}{|c|}{\bf $Z=112$ nuclei} \\ \hline
$^{292}$112 & 3.94 & 0.30 &      &                     & 3.64 &  (0.48,27$^{\circ}$) \\
$^{290}$112 & 3.48 & 0.31 &      &                     & 3.44 &  (0.47,26$^{\circ}$) \\
$^{288}$112 & 3.07 & 0.31 &      &                     &      &                      \\
$^{286}$112 & 2.90 & 0.32 &      &                     &      &                      \\
$^{284}$112 & 2.65 & 0.32 &      &                     &      &                      \\
$^{282}$112 & 2.35 & 0.41 &      &                     &      &                      \\
$^{280}$112 & 2.70 & 0.43 &      &                     &      &                      \\
$^{278}$112 & 3.06 & 0.44 &      &                     &      &                      \\
$^{276}$112 & 3.34 & 0.44 &      &                     &      &                      \\ \hline
\end{tabular}
\end{center}
\end{table}

\begin{table}[ptb]
\caption{The same as in Table\ \ref{tab5} but only for
nuclei where experimental estimates of inner fission
barrier heights exist. The results of calculations with
the DD-PC1 and DD-ME2 parametrizations are presented.
\label{tab7}}
\begin{center}
\renewcommand{\arraystretch}{1.2} 
\begin{tabular}{|c|c|c|c|c|c|c|}
\hline
Nucleus & $B_f^{Ax}$ & $\beta_2$ & $B_f^{X}$ & ($\beta_2$,$\gamma$) & $B_f^{Tr-B}$ & ($\beta_2$,$\gamma$)\\ \hline
\multicolumn{7}{|c|}{\bf DD-PC1 parametrization} \\ \hline
$^{294}$116 & 3.64 & 0.31 &  3.47*   & (0.32,10$^{\circ}$)  & 2.58 & (0.48,27$^{\circ}$) \\
$^{292}$116 & 3.63 & 0.32 &  3.55*   &  (0.32,5$^{\circ}$)  & 2.83 & (0.49,26$^{\circ}$) \\
$^{290}$114 & 3.42 & 0.34 &          &                      &      &                     \\
$^{288}$114 & 3.38 & 0.32 &          &                      &      &                     \\
$^{286}$112 & 3.02 & 0.31 &          &                      &      &                     \\
$^{284}$112 & 2.75 & 0.31 &          &                      &      &                     \\ \hline
\multicolumn{7}{|c|}{\bf DD-ME2 parametrization} \\ \hline
$^{294}$116 & 5.36 & 0.32 &  5.20*   & (0.32,8$^{\circ}$)   & 4.45 & (0.48,25$^{\circ}$) \\
$^{292}$116 & 4.77 & 0.32 &  4.68*   & (0.32,5$^{\circ}$)   & 4.24 & (0.48,27$^{\circ}$) \\
$^{290}$114 & 4.33 & 0.34 &          &                      &      &                     \\
$^{288}$114 & 4.18 & 0.32 &          &                      &      &                     \\
$^{286}$112 & 3.62 & 0.31 &          &                      &      &                     \\
$^{284}$112 & 3.12 & 0.31 &          &                      &      &                     \\ \hline
\end{tabular}
\end{center}
\end{table}

 It is interesting to compare the current results with the ones obtained in other
models. The results of Skyrme DFT calculations of Ref.\ \cite{SDN.06} for the
$N=184$ isotones show that the impact of triaxiality on the inner fission barrier
is small for $Z=112$, but increases with increasing Z (see Fig.\ 4 in Ref.\ \cite{SDN.06}).
The lowering of the inner fission barrier due to triaxiality is around 2 MeV for $Z=120$ and
exceeds 3 MeV for $Z=126$.  In the ETFSI model calculations \cite{DPT.00} the inner fission
barriers are lowered due to triaxiality in the nuclei $(Z=112, N=182)$ and $(Z=114, N=184)$
by 0.5 and 1.1 MeV, respectively. In the macroscopic+microscopic calculations
of Ref.\ \cite{KJS.10} the largest reduction of the inner barrier height due to
triaxiality is about 2 MeV and it appears in the region around $Z\approx 122$,
$N \approx 180$ (see Fig.\ 2 in Ref.\ \cite{KJS.10}).

  As discussed in Ref.\ \cite{AAR.10} on the example of actinide nuclei, the
reduction of the inner fission barrier height due to triaxiality is caused by the
level densities in the vicinity of the Fermi level which are lower at triaxial
shape as compared with axial one. The different location of the ``magic'' shell
gaps in superheavy nuclei in the macroscopic+microscopic model (at $Z=114$, $N=184$),
in Skyrme DFT (predominantly at $Z=126$, $N=184$) and in CDFT (at $Z=120$, $N=172$)
results in different deformed single-particle structures
at the deformations typical for the saddles of the inner fission barrier. This
is one of the sources of the differences in the predictions of different models.

 We have to keep in mind, however, that we use in this investigation a seniority
zero pairing force with a fixed cutoff energy of $E_{\rm cutoff}=120$ MeV.
It has been shown in Ref.~\cite{KALR.10} that pairing correlations play an important
role for the calculation of the fission barriers and that even if the
strength of the pairing force is adjusted to experimental gap parameters at the ground
state, the range of the pairing force has an influence on the height of the barrier.
This means for zero range forces and for the seniority zero forces that the barriers
depend on the cutoff energy. It also has been shown in Ref.~\cite{KALR.10} that the
parameter set DD-ME2 in connection with the finite range Gogny force D1S in the
pairing channel produces in axial symmetric calculations in most of the actinide nuclei
inner barriers which are too high as compared with experiment (see Fig. 9 of Ref.~\cite{KALR.10})
and that it produces for the superheavy elements with $Z=$ 112, 114, and 116,
where experimental estimates are available~\cite{IOZ.02} in axially symmetric calculations
barriers which are in rather good agreement with those values (see Fig. 10 of Ref.~\cite{KALR.10}).
Of course, so far, there exist no relativistic triaxial calculations with the finite range
Gogny force in the pairing channel and full Gogny calculations~\cite{DGGL.06} are hard
to compare because of the different spin-orbit force used in this model. We have,
however to keep in mind, that triaxiality reduces the barriers in the actinides, but
not in the superheavy elements with $Z=$ 112, 114, and 116. Therefore, at the
moment, we can only conclude that for a final comparison with the experimental
data we have to wait for full RHB calculations or at least RMF+BCS calculations
with the finite range Gogny force in the pairing channel. Investigations in
this direction are in progress.

\begin{figure}[ht]
\includegraphics[angle=0,width=8.0cm]{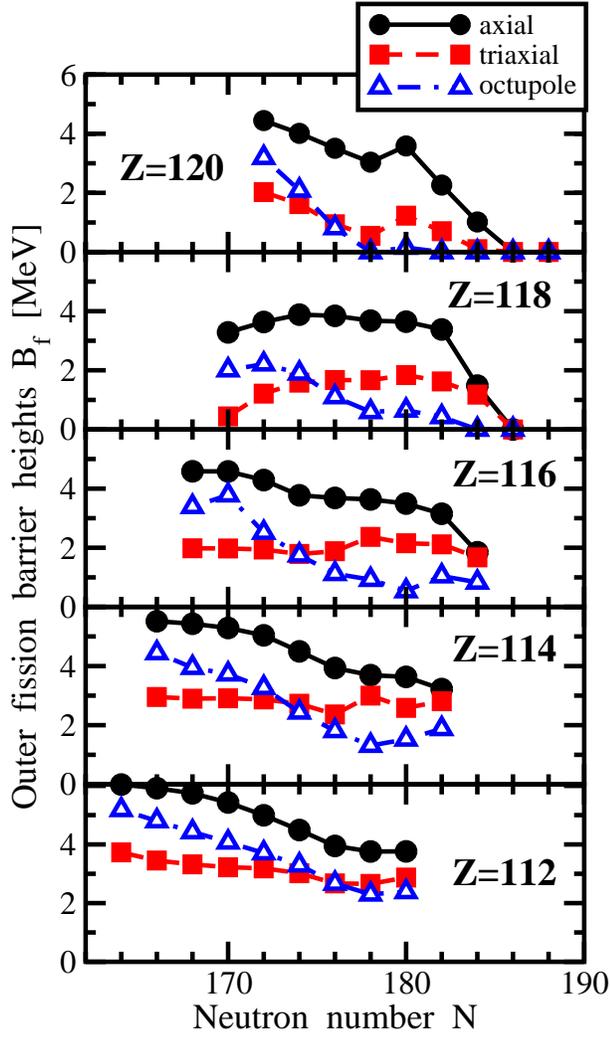}
\caption{(Color online) The heights of outer fission barriers of
even-even $Z=112-120$ nuclei relative to spherical/weakly(normal)-deformed
minima as a function of neutron number. The results of calculations of type A 
(axial), T (triaxial) and O (octupole) obtained with the NL3*
parametrization are presented.}
\label{OFB}
\end{figure}

It is well known that reflection-asymmetric (octupole deformed) shapes become
important at the second fission barrier and beyond (see Refs.\ \cite{SBDN.09,MSI.09}
and references therein). Our calculations indicate that triaxiality can play
a similarly important role at the outer fission barriers. Usually, the triaxiality
of these barriers is not mentioned in the publications. To our knowledge, it is only
discussed in Ref.\ \cite{DPB.07} that non-axial degrees of freedom

play an important role in the description of outer fission barriers of actinides.
Fig.\ \ref{OFB} compares the energies of outer barrier saddle points as obtained in
axial reflection symmetric (A), triaxial (T) and axial reflection asymmetric 
[octupole deformed] (O) calculations. One can see that the inclusion of triaxiality or
octupole deformation always lowers the outer fission barrier. The underlying shell
structure clearly defines which of the saddle points (triaxial or octupole deformed) 
is lower in energy. For example, the lowest saddle point is obtained in triaxial calculations
in proton-rich nuclei ($N < 174$). On the contrary, the lowest saddle point
is obtained in octupole deformed calculations in neutron-rich nuclei ($N > 174$).
Note that the decrease of the saddle point energy by octupole deformation
or triaxiality reaches 3 MeV in some nuclei. Thus, one can conclude
that due to the structure of the PES in the fission path valley, we observe in the
superheavy region an opposite situation as compared to actinide nuclei where the
triaxiality has no impact  on the outer fission barriers. Our results also suggest 
that in some superheavy nuclei [mostly in the nuclei where the deformation 
energy curves of the A and O calculations are similar in energy] the 
combination of two deformations (triaxiality 
and odd-multipole deformations) may be important in the definition of the fission path 
for $\beta_2 \geq 0.5$. The CDFT calculations with both deformations included are 
at present not yet possible, but require further investigations.

\begin{figure}[ht]
\includegraphics[angle=0,width=8.0cm]{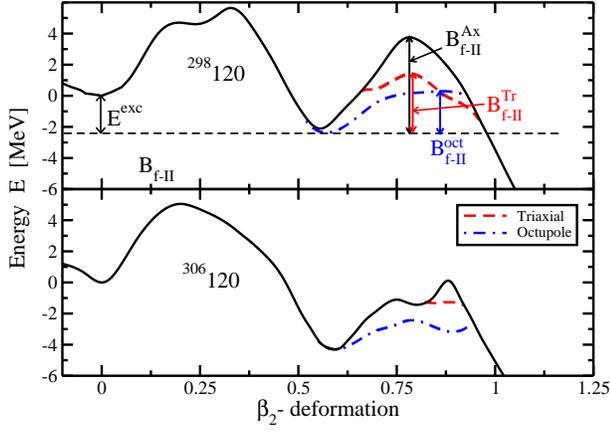}
\caption{(Color online) The same as in Fig.\ \ref{pot} but for the nuclei
$^{298}$120 and $^{306}$120. Note that for simplicity only
the axial solution (A) is shown at $\beta_2 < 0.5$.}
\label{oct-second}
\end{figure}

Fig.\ \ref{oct-second} shows that some nuclei are superdeformed in the ground
state. A summary for such nuclei is given in Table \ref{SD-summary}. Whether
these states are stable or metastable should be defined by the height and the width
of the outer fission barrier. The current calculations show that in many nuclei
this barrier is appreciable in the axial calculations of type A. However, the inclusion of
triaxial or octupole deformation decreases this barrier substantially so it
is around 2 MeV in the majority of the nuclei. This low barrier would translate
into a high penetration probability for spontaneous fission, such that most likely these
superdeformed states are metastable. Calculations of the spontaneous fission
half-lives from spherical/weakly-deformed and superdeformed minima are needed in order
to decide which of these minima is more stable against fission. Table \ref{SD-summary}
shows that the saddle point obtained in octupole deformed calculations (O) is the lowest
in energy in most of the cases. Only in proton-rich $^{294}$120 and $^{290,292}$118
nuclei, the saddle point of triaxial calculations (T) is the lowest.

A superdeformed minimum exists also in the doubly magic nucleus
$^{292}$120 at a low excitation energy of approximately 0.6 MeV (see Fig.\
\ref{pot}). This is definitely not an artifact of the model under consideration, since similar
minimum exists also in axial relativistic Hartree-Bogoliubov calculations with the DD-ME2
parametrization with Gogny D1S forces in the pairing channel (see Fig.\ 8 in Ref.\
\cite{KALR.10}). Its calculated excitation energy depends on the actual strength of
pairing and varies between 0 and 2 MeV.

It is interesting to compare our results for the structure of the outer fission
barriers with those obtained in the axial RMF+BCS model (with and without reflection
symmetry) of Ref.\ \cite{BBM.04} which employed the NL3 and NL-Z2 parametrizations. 
Similar to our calculations, the superdeformed minima exist in the calculations of 
Ref.\ \cite{BBM.04} (see Fig.\ 5 in this reference) without octupole deformation. 
However, in the calculations  of Ref.\ \cite{BBM.04} with the NL-Z2 parametrization, 
the heights of outer fission barriers with respect to the SD minimum are lower by approximately 
2 MeV as compared to our calculations. As a consequence, the inclusion of octupole deformation
completely eliminates the outer fission barriers leading to results contradicting
ours. On the other hand, the results of the calculations of Ref.\ \cite{BBM.04}
(see Fig.\ 6 in this reference) for a few selected nuclei based on the NL3 parametrization
are very similar to ours, namely, the ground state is superdeformed and the outer fission
barrier has a height of approximately 2.5 MeV. This result is not surprising considering
that the NL3* parametrization is very similar to NL3 \cite{NL3*}. However,
the results of the calculations for a few selected $Z=112, 114$ and 116 nuclei shown in
Figs.\ \ref{pot3a} and \ref{pot3b} reveal that the outer fission barriers survive
in the presence of octupole and triaxial deformation not only in the NL3* parametrization
but also in the DD-ME2 and DD-PC1 parametrizations. On average, the height of outer
fission barrier in these nuclei is around 2 MeV. On the contrary, the calculations
with the NL-Z2 parametrization in Ref.\ \cite{BBM.04} show that octupole deformation
kills the outer fission barriers in these nuclei (see Fig.\ 5 in this reference).

\begin{figure}[ht]
\includegraphics[angle=0,width=8.0cm]{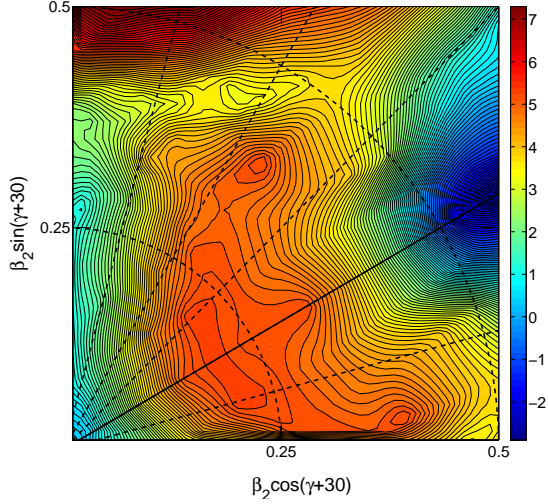}
\caption{(Color online) The same as in Fig.\ \ref{pes-2D-6panel} but for 
the nucleus $^{302}$120. In order to show the potential energy surface 
between the axial $\beta\sim 0.2$ and triaxial ($\beta_2\sim 0.40, \gamma\sim 25^{\circ}$) 
hills in greater details the energy difference between two neighboring 
equipotential lines is set to 0.1 MeV and only one fourth of deformation
plane of Fig.\ \ref{pes-2D-6panel} is shown.}
\label{302-valley}
\end{figure}

\begin{table}[ptb]
\caption{{A summary of the nuclei in which the superdeformed
minimum is the lowest in energy. The energies of outer fission barrier
saddle points with respect to this minimum, as obtained in axial
($B_{f-II}^{A}$), triaxial ($B_{f-II}^{T}$) and octupole ($B_{f-II}^{O}$)
deformed calculations, are shown in columns 3, 4, and 5, respectively.
For each nucleus, the outer saddle point with lowest energy is shown in 
bold. The excitation energies $E^{exc}$ of spherical/weakly(normal)-deformed 
minima with respect to the superdeformed minima are shown in column 2. The 
graphical explanation of these quantities is also given in Fig.\ 
\ref{oct-second}.}
\label{SD-summary}}
\begin{center}
\begin{tabular}{|c|c|c|c|c|}
\hline
Nucleus & $E^{exc}$   & $B_{f-II}^{A}$ & $B_{f-II}^{T}$ & $B_{f-II}^{O}$\\ \hline
      1 &           2 & 3               & 4               & 5 \\ \hline
\multicolumn{5}{|c|}{\bf The $Z=120$ nuclei} \\ \hline
$^{308}120$ & 4.84    & 4.52            & 2.75             & {\bf 1.62}   \\   \hline
$^{306}120$ & 4.26    & 4.36            & 3.00             & {\bf 1.82}   \\ \hline
$^{304}120$ & 3.48    & 4.49            & 3.57             & {\bf 2.07}   \\ \hline
$^{302}120$ & 2.89    & 5.30            & 3.65             & {\bf 2.55}   \\ \hline
$^{300}120$ & 2.60    & 6.18            & 3.82             & {\bf 2.61}   \\ \hline
$^{298}120$ & 1.99    & 5.71            & 3.38             & {\bf 2.25}   \\ \hline
$^{296}120$ & 1.72    & 4.87            & 2.94             & {\bf 2.52}   \\ \hline
$^{294}120$ & 0.62    & 4.63            & {\bf 2.24}       &  2.71        \\ \hline
\multicolumn{5}{|c|}{\bf The $Z=118$ nuclei} \\ \hline
$^{304}118$ & 2.68    & 2.41            & 2.14             &  {\bf 1.31}  \\ \hline
$^{302}118$ & 2.92    & 3.58            & 2.42             &  {\bf 1.42}  \\ \hline
$^{300}118$ & 1.78    & 5.16            & 3.40             &  {\bf 2.20}  \\ \hline
$^{298}118$ & 1.70    & 5.34            & 3.54             &  {\bf 2.34}  \\ \hline
$^{296}118$ & 1.48    & 5.17            & 3.15             &  {\bf 2.26}  \\ \hline
$^{294}118$ & 1.15    & 4.99            & 2.82             &  {\bf 2.35}  \\ \hline
$^{292}118$ & 0.69    & 4.58            & {\bf 2.28}       &  2.59        \\ \hline
$^{290}118$ & 0.40    & 4.03            & {\bf 1.61}       &  2.60        \\ \hline
\multicolumn{5}{|c|}{\bf The $Z=116$ nuclei} \\ \hline
$^{300}116$ & 0.92    & 2.76            & 2.60             &  {\bf 1.75}  \\ \hline
$^{298}116$ & 0.83    & 3.98            & 2.95             &  {\bf 1.88}  \\ \hline
\end{tabular}
\end{center}
\end{table}

\section{Conclusions}
\label{conc}

We have carried out first systematic investigations of fission barriers in
even-even superheavy nuclei with $Z=112-120$ within covariant density 
functional theory including triaxial shapes with D2-symmetry and octupole 
shapes with axial symmetry. Three different classes of models with the 
state-of-the-art parameterizations NL3*, DD-ME2 and DD-PC1 were used in the 
calculations. Pairing correlations are taken into account in the BCS 
approximation using seniority pairing forces adjusted to empirical values 
of the gap parameters. The following conclusions have been obtained:

\begin{itemize}

\item
 The low-$Z$ and low-$N$ nuclei in this region are characterized by axially
symmetric inner fission barriers. The increase of the particle numbers leads
to a softening of the potential energy surfaces in the triaxial plane.
As a result, several competing fission pathes in the region of inner fission
barrier emerge in some of the nuclei. Their importance in spontaneous fission
can be defined only by taking into account the fission dynamics more seriously.
However, the results of the calculations within the macroscopic+microscopic 
method \cite{GSPS.99} suggests that the 'Tr-B' path (and even maybe 'Tr-A' 
one) may not be so important since they are substantially longer as compared
with the axially symmetric path because it leads to a significant reduction 
of the penetration probability.

\item
Triaxiality lowers the outer fission barriers by 1.5-3 MeV in reflection
symmetric calculations. In many nuclei the lowering due to triaxiality
is even more important than the one due to octupole deformation, which
known to be important for the outer fission barriers and beyond from previous
calculations (see Ref.\ \cite{BBM.04} and references therein). The 
underlying shell structure clearly defines that the triaxial [octupole] 
saddle is lower in energy for proton-rich nuclei with $N < 174$ [neutron-rich 
nuclei with $N > 174$].

\item
On average, inner and outer fission barriers obtained for the NL3* and
DD-PC1 parametrizations are similar. On the contrary, the DD-ME2 parametrization
produces barriers which are by 1-1.5 MeV higher than the ones obtained
with NL3* and DD-PC1.

\item
The superdeformed minimum is the lowest in energy in some of these nuclei.
In the present calculations, the outer fission barriers with respect 
to these minima are about 2 MeV high. Both minima and barriers are present 
in all three classes of the CDFT models. It has to be investigated, 
however, if these superdeformed minima are stable with respects to
more general deformations not taken into account so far.

\end{itemize}

   The comparison of our results with those of non-relativistic models
clearly shows that CDFT predictions for the heights of inner fission
barriers in the superheavy region with a seniority zero pairing force
still remain on the lower end among nuclear structure models used so far.
Only axially symmetric calculations with the finite range Gogny force
D1S in the pairing channel can reproduce the estimates of inner fission barrier
heights of Ref.\ \cite{IOZ.02}. Considering the uncertainties of these 
estimates, the investigation of other experimental observables that strongly 
depend on fission barrier
heights, especially fission half-lives and $\beta$- or electron capture
delayed fission, is needed. The work in this direction is in progress.

\section{Acknowledgements}

  This work has been supported by the U.S. Department of Energy under
the grant DE-FG02-07ER41459, by a travel grant to JUSTIPEN (Japan-US
Theory Institute for Physics with Exotic Nuclei) under U.\ S. Department
of Energy Grant DE-FG02-06ER41407 and by the DFG cluster of excellence
\textquotedblleft Origin and Structure of the Universe
\textquotedblright\ (www.universe-cluster.de).


\begin{thebibliography}{100}

\bibitem{AAR.10}
H. Abusara, A.~V. Afanasjev, and P. Ring, Phys. Rev. {\bf C82},  044303
  (2010).

\bibitem{SBDN.09}
A. Staszczak, A. Baran, J. Dobaczewski, and W. Nazarewicz, Phys. Rev. {\bf
  C80},  014309  (2009).

\bibitem{WERP.02}
M. Warda, J.~L. Egido, L.~M. Robledo, and K. Pomorski, Phys. Rev. {\bf C66},
  014310  (2002).

\bibitem{MSI.04}
P. M{\"o}ller, A.~J. Sierk, and A. Iwamoto, Phys. Rev. Lett. {\bf 92},  072501
  (2004).

\bibitem{SK.06}
A. Sobiczewski and M. Kowal, Phys. Scripta {\bf T125},  68  (2006).

\bibitem{DPB.07}
J. Dobrowolski, K. Pomorski, and J. Bartel, Phys. Rev. {\bf C75},  024613
  (2007).

\bibitem{MSI.09}
P. M{\"o}ller, A.~J. Sierk, T. Ichikawa, A. Iwamoto, R. Bengtsson, H.
  Uhrenholt, and S. {\AA}berg, Phys. Rev. {\bf C79},  064304  (2009).

\bibitem{DNPB.09}
A. Dobrowolski, B. Nerlo-Pomorska, K. Pomorski, and J. Bartel, Acta Phys.\
  Polonica {\bf B40},  705  (2009).

\bibitem{KJS.10}
M. Kowal, P. Jacimowicz, and A. Sobiczewski, Phys. Rev. {\bf C82},  014303
  (2010).

\bibitem{DPT.00}
A.~K. Dutta, J.~M. Pearson, and F. Tondeur, Phys. Rev. {\bf C61},  054303
  (2000).

\bibitem{BRRMG.98}
M. Bender, K. Rutz, P.-G. Reinhard, J.~A. Maruhn, and W. Greiner, Phys. Rev.
  {\bf C58},  2126  (1998).

\bibitem{BQS.04}
L. Bonneau, P. Quentin, and D. Samsoen, Eur. Phys. J. {\bf A21},  391  (2004).

\bibitem{SDN.06}
A. Staszczak, J. Dobaczewski, and W. Nazarewicz, Acta Phys. Polonica B {\bf
  38},  1589  (2007).

\bibitem{DGGL.06}
J.-P. Delaroche, M. Girod, H. Goutte, and J. Libert, Nucl. Phys. {\bf A771},
  103  (2006).

\bibitem{WAR.09}
M. Warda, Eur. Phys. J. {\bf A42},  605  (2009).

\bibitem{RSRG.10}
R. Rodr{\'i}guez-Guzm{\'a}n, P. Sarriguren, L.~M. Robledo, and J.~E.
  Garcia-Ramos, Phys. Rev. {\bf C81},  024310  (2010).

\bibitem{LNVR.10}
Z.~P. Li, T. Nik\v{s}i\'{c}, D. Vretenar, P. Ring, and J. Meng, Phys. Rev. {\bf
  C81},  064321  (2010).

\bibitem{RAALNV.11}
P. Ring, H. Abusara, A.~V. Afanasjev, G.~A. Lalazissis, T. Nik\v{s}i\'{c}, and
  D. Vretenar, Int. J. Mod. Phys. {\bf E20},  235  (2011).

\bibitem{KS.65}
W. Kohn and L.~J. Sham, Phys. Rev. {\bf 137},  A1697  (1965).

\bibitem{KS.65a}
W. Kohn and L.~J. Sham, Phys. Rev. {\bf 140},  A1133  (1965).

\bibitem{LNP.641}
{\em Extended Density Functionals in Nuclear Structure Physics}, Vol.~641 of
  {\em Lecture Notes in Physics}, edited by G.~A. Lalazissis, P. Ring, and D.
  Vretenar (Springer-Verlag, Heidelberg, 2004), .

\bibitem{CFG.92}
T.~D. Cohen, R.~J. Furnstahl, and K.Griegel, Phys. Rev. {\bf C45},  1881
  (1992).

\bibitem{KR.89}
W. Koepf and P. Ring, Nucl. Phys. {\bf A493},  61  (1989).

\bibitem{AA.10}
A.~V. Afanasjev and H. Abusara, Phys. Rev. {\bf C81},  014309  (2010).

\bibitem{HR.88}
U. Hofmann and P. Ring, Phys. Lett. {\bf B214},  307  (1988).

\bibitem{AR.00} A.~V. Afanasjev and P. Ring,
Phys.\ Rev. C {\bf 62}, 031302(R) (2000).


\bibitem{TO-rot} A.\ V.\ Afanasjev and H.\ Abusara,
Phys.\ Rev. C {\bf 82} 034329 (2010).

\bibitem{BT.92}
R. Brockmann and H. Toki, Phys. Rev. Lett. {\bf 68},  3408  (1992).

\bibitem{HKL.01}
F. Hofmann, C.~M. Keil, and H. Lenske, Phys. Rev. {\bf C64},  034314  (2001).

\bibitem{SOA.05}
M. Serra, T. Otsuka, Y. Akaishi, P. Ring, and S. Hirose, Prog. Theor. Phys.
  {\bf 113},  1009  (2005).

\bibitem{HSR.07}
S. Hirose, M. Serra, P. Ring, T. Otsuka, and Y. Akaishi, Phys. Rev. {\bf C75},
  024301  (2007).

\bibitem{DD-ME1}
T. Nik{\v{s}}i{\'{c}}, D. Vretenar, P. Finelli, and P. Ring, Phys. Rev. {\bf
  C66},  024306  (2002).

\bibitem{DD-ME2}
G.~A. Lalazissis, T. Nik{\v{s}}i{\'{c}}, D. Vretenar, and P. Ring, Phys. Rev.
  {\bf C71},  024312  (2005).

\bibitem{DD-PC1}
T. Nik\v{s}i\'{c}, D. Vretenar, and P. Ring, Phys. Rev. {\bf C78},  034318
  (2008).

\bibitem{DD-ME3}
X. Roca-Maza, X. Vi{\~n}as, M. Centelles, P. Ring, and P. Schuck, Phys. Rev. C
  {\bf 84},  054309  (2011).

\bibitem{VALR.05}
D. Vretenar, A.~V. Afanasjev, G.~A. Lalazissis, and P. Ring, Phys. Rep. {\bf
  409},  101  (2005).

\bibitem{NVR.11}
T. Nik\v{s}i\'{c}, D. Vretenar, and P. Ring, Prog. Part. Nucl. Phys. {\bf 66},
  519  (2011).

\bibitem{NL3*}
G.~A. Lalazissis, S. Karatzikos, R. Fossion, D. Pe{\~n}a~Arteaga, A.~V.
  Afanasjev, and P. Ring, Phys. Lett. {\bf B671},  36  (2009).

\bibitem{TW.99}
S. Typel and H.~H. Wolter, Nucl. Phys. {\bf A656},  331  (1999).

\bibitem{GRT.90}
Y.~K. Gambhir, P. Ring, and A. Thimet, Ann. Phys. (N.Y.) {\bf 198},  132
  (1990).

\bibitem{Wal.74}
J.~D. Walecka, Ann. Phys. (N.Y.) {\bf 83},  491  (1974).

\bibitem{SW.86}
B.~D. Serot and J.~D. Walecka, Adv. Nucl. Phys. {\bf 16},  1  (1986).

\bibitem{BB.77}
J. Boguta and A.~R. Bodmer, Nucl. Phys. {\bf A292},  413  (1977).

\bibitem{NL3}
G.~A. Lalazissis, J. K{\"o}nig, and P. Ring, Phys. Rev. {\bf C55},  540
  (1997).

\bibitem{NHM.92}
A.~A. Nikolaus, T. Hoch, and D. Madland, Phys. Rev. {\bf C46},  1757  (1992).

\bibitem{KR.88}
W. Koepf and P. Ring, Phys. Lett. {\bf B212},  397  (1988).

\bibitem{RS.80}
P. Ring and P. Schuck, {\em The Nuclear Many-Body Problem} (Springer-Verlag,
  Berlin, 1980).

\bibitem{MN.92}
P. M{\"o}ller and J. Nix, Nucl. Phys. {\bf A536},  20  (1992).

\bibitem{RIPL-2} RIPL-2 stands for reference input parameter library of
International Atomic Energy Agency located at http://www-nds.iaea.org/ripl2/,
which for actinides is based on Ref.\protect\cite{M.98}.

\bibitem{DMS.80}
J. Dudek, A. Majhofer, and J. Skalski, J. Phys. {\bf G6},  447  (1980).

\bibitem{MPS.01}
I. Muntian, Z. Patyk, and A. Sobicziewski, Acta Physica Polonica. {\bf B32},
  691  (2001).

\bibitem{BBM.04}
T. B{\"u}rvenich, M. Bender, J.~A. Maruhn, and P.-G. Reinhard, Phys. Rev. {\bf
  C69},  014307  (2004).

\bibitem{BRRM.00}
M. Bender, K. Rutz, P.-G. Reinhard, and J.~A. Maruhn, Euro. Phys. J. {\bf A8},
  59  (2000).

\bibitem{SGP.05}
M. Samyn, S. Goriely, and J.~M. Pearson, Phys. Rev. {\bf C72},  044316  (2005).

\bibitem{MNK.97}
P. M{\"o}ller, J.~R. Nix, and K.~L. Kratz, At. Data Nucl. Data Tables {\bf 66},
   131  (1997).

\bibitem{SDN.07}
A. Staszczak, J. Dobaczewski, and W. Nazarewicz, Int. J. Mod. Phys. {\bf E16},
  310  (2007).

\bibitem{KALR.10}
S. Karatzikos, A.~V. Afanasjev, G.~A. Lalazissis, and P. Ring, Phys. Lett. {\bf
  B689},  72  (2010).

\bibitem{BAM.08}
A. Baran, A. Bulgac, M.~M. Forbes, G. Hagen, W. Nazarewicz, N. Schunck, and
  M.~V. Stoitsov, Phys. Rev. {\bf C78},  014318  (2008).

\bibitem{SSBN.10}
A. Staszack, M. Stoitsov, A. Baran, and W. Nazarewicz, Eur. Phys. J. {\bf A46},
   85  (2010).

\bibitem{RGL.97} P.\ Ring, Y.\ K.\ Gambhir, and G.\ A.\ Lalazissis,
Comp.\ Phys.\ Comm. {\bf 105}, 77 (1997).

\bibitem{AKR.99}
A.~V. Afanasjev, J. K{\"o}nig, and P. Ring, Phys. Rev. {\bf C60},  051303(R)
  (1999).

\bibitem{ARK.00}
A.~V. Afanasjev, P. Ring, and J. K{\"o}nig, Nucl. Phys. {\bf A676},  196
  (2000).

\bibitem{ERE.07} W.\ E, W.\ Ren, and E.\ Vanden-Eijnden, J.\ Chem.\ 
Phys. {\bf 126}, 164103 (2007).

\bibitem{STH.08} D.\ Sheppard, R.\ Terrell and G.\ Henkelman,
J.\ Chem.\ Phys. {\bf 128}, 134106 (2008).

\bibitem{MM.11} D.\ G.\ Madland and P.\ M{\"o}ller, {\it Los Alamos National
Laboratory unclassified report}, LA-UR-11-11447 (2011).

\bibitem{AKF.03}
A.~V. Afanasjev, T.~L. Khoo, S. Frauendorf, G.~A. Lalazissis, and I. Ahmad,
  Phys. Rev. {\bf C67},  024309  (2003).

\bibitem{GSPS.99} R.\ A.\ Gherghescu, J.\ Skalski, Z.\ Patyk, and
                  A.\ Sobiczewski, {\it Nucl.\ Phys.} {\bf A651} (1999) 237.


\bibitem{IOZ.02}
M.~G. Itkis, Y.~T. Oganessian, and V.~I. Zagrebaev, Phys. Rev. {\bf C65},
  044602  (2002).

\bibitem{M.98} V.\ M.\ Maslov. {\it RIPL-1 Handbook.} TEXDOC-000, IAEA,
               Vienna, 1998, Ch.\ 5.

\end{thebibliography}
\end{document}